\documentclass{IEEEtran}
\usepackage{cite}
\usepackage{amsmath,amssymb,amsfonts}
\usepackage{algorithmic}
\usepackage{graphicx}
\usepackage{array}
\usepackage{textcomp}
\usepackage{threeparttable}
\usepackage{comment}
\usepackage{multicol, blindtext,xcolor}
\definecolor{darkgreen}{rgb}{0.0, 0.5, 0.0}
\def\BibTeX{{\rm B\kern-.05em{\sc i\kern-.025em b}\kern-.08em
    T\kern-.1667em\lower.7ex\hbox{E}\kern-.125emX}}
\begin{document}
\title{Arbitrary Diffraction Engineering with\\Multilayered Multielement Metagratings }
\author{Oshri Rabinovich, \IEEEmembership{Student Member, IEEE}, and Ariel Epstein, \IEEEmembership{Member, IEEE}
\thanks{The authors are with the Andrew and Erna Viterbi Faculty of Electrical Engineering, Technion - Israel Institute of Technology, Haifa 3200003, Israel (e-mail: oshrir@technion.ac.il; epsteina@ee.technion.ac.il).}
\thanks{Manuscript received XX,YY,2019; revised XX,YY, 2019.}}

\markboth{}%
{Rabinovich and Epstein}

\maketitle

\begin{abstract}

We theoretically formulate and experimentally demonstrate an analytical formalism for the design of printed circuit board (PCB) metagratings (MGs) exercising individual control over the \emph{amplitude} and \emph{phase} of numerous diffracted modes, in both \emph{reflection} and \emph{transmission}. Lately, these periodic arrangements of subwavelength polarizable particles (meta-atoms) were shown to deflect an incoming plane wave to prescribed angles with very high efficiencies, despite their sparsity with respect to conventional metasurfaces. Nonetheless, most reported MGs were designed based on full-wave optimization of the meta-atoms, with the scarce analytical schemes leading directly to realizable devices were restricted to single-layer reflecting structures, controlling only the partition of power.
In this paper, we present an analytical model for plane-wave interaction with a general multilayered multielement MG, composed of an arbitrary number of meta-atoms distributed across an arbitrary number of layers in a given stratified media configuration. For a desired (forward and backward) diffraction pattern, we formulate suitable constraints, identify the required number of degrees of freedom, and correspondingly set them to yield a detailed MG configuration implementing the prescribed functionality; no full-wave optimization is involved. To verify and demonstrate the versatility of this systematic approach, 
fabrication-ready multilayer PCB MGs for perfect anomalous refraction and non-local focusing are synthesized, fabricated, and experimentally characterized, for the first time to the best of our knowledge, indicating very good correspondence with theoretical predictions. Importantly, the formalism also accounts for realistic conduction losses, critical to obtain reliable efficient designs. 
This appealing semianalytical methodology is expected to accelerate the development of MGs and extend the relevant range of applications, yielding practical MG designs on demand for arbitrary beam manipulation.


\end{abstract}


\section{Introduction}
\label{Sec:Introduction}
\IEEEPARstart{O}{ne} of the earliest roles of optics was to manipulate the
trajectories of incident plane waves. Even today, lenses, mirrors, and beam splitters, still form the basic building blocks of many optical systems \cite{born1999principles,goodman2005introduction,yariv2006photonics}; such devices are of utmost importance also for applications at microwave frequencies, e.g., for radar cross section (RCS) reduction, or for directive lens and reflector antennas \cite{balanis2005antenna,balanis2012advanced,knott2004radar}. It is not surprising, thus, that much effort has been devoted in the past decades to reduce the size and enhance the performance of such devices. 

One methodology to deflect a beam via a compact formation is to use diffraction gratings \cite{fano1941theory,loewen1997diffraction}. These planar periodic structures obey the Floquet-Bloch (FB) theorem, implying that an incident plane wave would be scattered into specific allowed directions, determined by the angle of incidence and the grating period. While these two parameters can be readily tuned to meet the required (anomalous) scattering angle, coupling to spurious diffraction modes would, in general, result in suboptimal efficiency. This is due to the fact that the possibility of scattering towards undesirable angles always exists; at the very least, the fundamental (0th-order) FB modes, corresponding to specular reflection and direct ray transmission, are always allowed to propagate. 

The way to control the modal coupling efficiencies themselves passes through a judicious design of the grating geometry, i.e. by engineering the profile of a dielectric slab or the layout of conductors, within a period \cite{fano1941theory,loewen1997diffraction}. Nonetheless, to the best of our knowledge, despite the vast research on this topic, no rigorous accurate formulation facilitating reliable arbitrary control over the coupling to numerous diffraction modes, both in reflection and transmission, is available to date. Instead, to realize high-efficiency diffraction
gratings, common methodologies combine intuitive approximate approaches with full-wave optimization of the grating physical structure \cite{perry1995high,hehl1999high}. Moreover, even these methods typically succeed in managing the coupling to only one or two propagating FB modes; thus, to synthesize more complicated large aperture wavefront manipulating devices, e.g. axicons or lenses, diffraction engineering must be employed separately at different sections along the surface, locally redirecting each incident ray. Such an approach inherently relies on the geometrical optics approximation, which generally limits controlled field manipulation in close proximity to the device aperture \cite{felsen1994radiation}.

Another option for achieving wavefront control with ultrathin devices is to harness recent advances in the field of metasurfaces \cite{bomzon2001pancharatnam,yu2011light,pfeiffer2013metamaterial,monticone2013full,selvanayagam2013discontinuous,pfeiffer2013millimeter,
pfeiffer2014efficient,kim2014optical,asadchy2015functional,epstein2016huygens,estakhri2016recent}. These surfaces, composed of closely-packed subwavelength polarizable particles (meta-atoms), were shown to be modelled well by equivalent (homogenized) boundary conditions, also known as the generalized sheet transition conditions (GSTCs) \cite{tretyakov2003analytical,kuester2003averaged}. In recent years, large body of work on these structures revealed their ability to efficiently manipulate electromagnetic fields in a variety of ways \cite{glybovski2016metasurfaces}. In particular, and in contrast to diffraction gratings, rigorous analytical solutions for perfect anomalous refraction, reflection, and beam splitting, were found, yielding the bianisotropic GSTCs required to accurately implement optimal steering of a given incident plane wave towards a prescribed angle \cite{wong2016reflectionless,epstein2016arbitrary,asadchy2016perfect,epstein2016synthesis,kwon2017perfect}. 

Although the homogenization approximation provides an efficient set of analytical tools for the synthesis of
metasurfaces with intricate functionalities, the final result of such macroscopic designs is an abstract continuous surface constituent distribution. To realize these metasurfaces in practice, this continuous distribution has to be discretized, and the physical structure of individual meta-atoms has to be optimized to implement the local response prescribed by the GSTCs at each point along the surface \cite{epstein2016huygens}. This combined macroscopic+microscopic metasurface synthesis methodology is associated with two problematic issues. First, the microscopic (meta-atom) design heavily relies on time-consuming full-wave simulations, forming one of the main factors impeding developments in the field \cite{pfeiffer2014bianisotropic,epstein2016cavity,lavigne2018susceptibility,chen2018theory}. Even though several analytical techniques have been devised to design multilayered meta-atoms, typically required for realizing bianisotropic metasurfaces \cite{epstein2016arbitrary,monticone2013full,pfeiffer2014bianisotropic}, these do not capture well all the relevant phenomena (near-field inter-layer coupling, conductor and dielectric losses); thus, resorting to full-wave optimization is inevitable in almost all cases. 

Second, while the transition between the continuous homogenized GSTCs and the final implementation with discrete meta-atoms has been demonstrated to work well on many occasions, the range of validity of the homogenization approximation is not clearly defined, especially for metasurfaces featuring highly-inhomogeneous polarizability distributions. For instance, the intercoupling between non-identical adjacent meta-atoms is typically not considered in the microscopic design scheme; it is also not a priori clear how crucial the discretization resolution to the successful implementation of the macroscopic design \cite{estakhri2016wave}, or how the meta-atom spatial dispersion would affect the global metasurface performance \cite{tretyakov2015metasurfaces}. Thus, overall, if the detailed design performance does not match the predictions of the homogenized model, one may encounter troubleshooting difficulties.

Finally, although it was shown that passive lossless bianisotropic metasufraces can be designed to implement any given field transformation that locally conserves the real power \cite{epstein2016arbitrary}, intentional introduction of suitable auxiliary evanescent fields is usually required to sustain this local power conservation \cite{epstein2016synthesis}. These surface waves are often nontrivial; thus, even though metasurfaces allow, in principal, arbitrary field manipulation, in the absence of a systematic procedure to conceive the proper auxiliary fields, only a handful of functionalities can be accurately realized \cite{epstein2016synthesis,epstein2017arbitrary,kwon2017perfect,kwon2018lossless,kwon2018lossless1}. While for some applications, using two bianisotropic metasurfaces in cascade offers a means to extend the range of implementable field transformations, such solutions require design and alignment of multiple devices, and pose certain constraints on the compactness of the compound system \cite{dorrah2018bianisotropic,raeker2019compound}.

\begin{figure*}[t]
\centering
\includegraphics[width=7.2in]{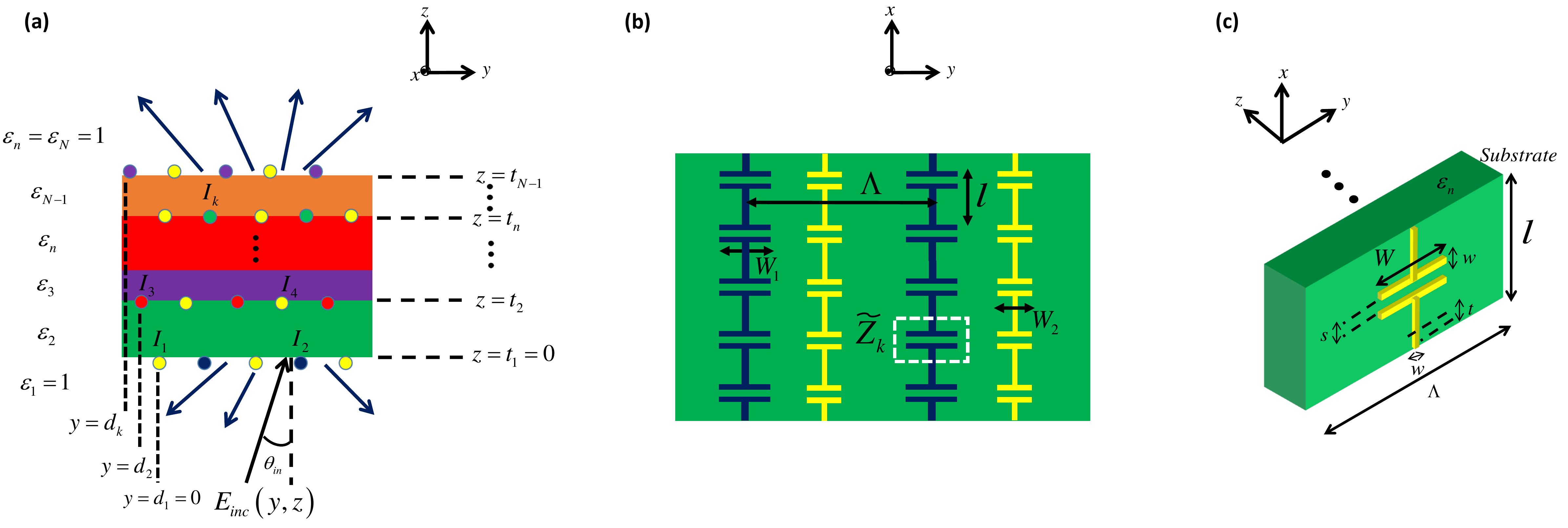}
\caption{Physical configuration of a multilayered multielement PCB MG.
(a)  Front view of the MG, with $K$ loaded strips distributed along the $N-1$ interfaces of the stratified dielectric media $\varepsilon_n$, with horizontal and vertical offsets $d_k$ and $h_k$, respectively, excited by a plane wave from below. (b)  Top view of one of the layer interfaces, consisting, for example, of two different loaded wires (two meta-atoms) per period $\Lambda$; the equivalent load impedance per unit length of the capacitive load of the $k$th wire is denoted by $\tilde{Z}_k$. 
(c) Trimetric view of a single meta-atom, featuring a printed capacitor of width $W$, repeating along the $x$ axis with periodicity $l\ll\lambda$. The trace width, trace separation, and copper thickness, are denoted by $w$, $s$, and $t$, respectively.}
\label{Fig:Config}
\end{figure*}

In this paper, we mitigate these issues by introducing a systematic rigorous methodology to design fabrication-ready multilayered multielement metagratings (MGs) for arbitrary diffraction engineering, without as much as a single simulation in a full-wave solver.
Based on the recently proposed idea of single- or dual-element MGs for perfect anomalous reflection, systematically put forward in \cite{ra2017meta,chalabi2017efficient}, we show that the (forward and backward) scattering to multiple FB modes can be simultaneously controlled, in both amplitude and phase, by a properly engineered distribution of multiple meta-atoms across multiple dielectric layers (Fig. \ref{Fig:Config}). We utilize capacitively-loaded wires as meta-atoms in our multilayered printed-circuit-board (PCB) device, extending our previously developed \cite{epstein2017unveiling,rabinovich2018analytical} (and experimentally verified \cite{rabinovich2019experimental}) detailed rigorous analytical model to relate the scattered fields to the individual wire positions and capacitor geometries. 
In the extended model, incorporating an arbitrary number of elements and an arbitrary stack of different dielectric layers, the full electromagnetic coupling between the meta-atoms (including multiple reflections and near field phenomena) is taken into account, allowing reliable resolution of the currents induced on the various wires in response to the incident fields, and subsequently the scattered ones. 

Matching the number of propagating FB modes to the number of different meta-atoms in a period\footnote{When dielectric laminate thicknesses are limited to a predetermined set of values, e.g., due to commercial availability, additional meta-atoms might be needed (see discussion in Sections \ref{subsec:passivity}, \ref{sec:results}, and the Appendix)}, we formulate a set of linear equations that coerce the induced currents to radiate FB modes with individually prescribed amplitude and phase. Subsequently, we devise an additional set of nonlinear constraints, guaranteeing that the required currents could be generated by loading the conducting wires with reactive (capacitive) elements, such that each meta-atom is passive and lossless. Resolving these two sets of linear and nonlinear equations while minimizing power dissipation in realsitic conducting wires, estimated analytically following\cite{epstein2017unveiling}, yields the complete conductor layout, i.e. the wire positions and the capacitor widths, to implement the desirable diffraction pattern. 
Astoundingly, without employing even a single full-wave optimization, the prescribed fabrication-ready MG geometry is shown to reliably couple the incident power to the various FB modes, exercising meticulous control over the \emph{complex} amplitudes of numerous modes. 

We demonstrate the unique capabilities of the developed multilayered mulielement MGs by designing two prototypical diffraction engineering devices: a 2-layer 4-element perfect anomalous refraction surface, realizing highly efficient wide-angle beam deflection; and a 4-layer 12-element non-local lens array, featuring diffraction limited focal spots. The theoretically predicted performance is verified with both full-wave simulations and experimental measurements, revealing excellent agreement for both prototypes. 
These case studies illustrate the great potential of the proposed devices for a wide variety of applications. Formed following an efficient bottom-up synthesis scheme, entirely avoiding full-wave optimization, they present a unique alternative to the time-consuming homogenization-based design of contemporary metasurfaces. Furthermore, as pointed out in previous work (e.g., \cite{wong2018perfect}), the evanescent modes required to sustain the desired functionality are inherently formed by the induced currents on the individual wires, avoiding the need to conceive and stipulate auxiliary fields as required for advanced metasurface designs \cite{epstein2016synthesis,epstein2017arbitrary,kwon2017perfect,kwon2018lossless,kwon2018lossless1}.

It should be noted that the field of beam-manipulating MGs has seen a drastic progress in the past couple of years; in particular, anomalous reflection or refraction using MGs has been demonstrated experimentally \cite{khaidarov2017asymmetric,yang2017freeform,paniagua2018metalens,sell2018ultra,fan2018perfect,wong2018perfect,wong2018binary,neder2019combined}. However, these reports addressed scenarios that allowed a limited number of propagating modes, and typically used a specifically tailored polarizable element, designed using full-wave optimization to meet the required performance. On the other hand, Popov \textit{et al.} have extended our analytical model \cite{rabinovich2018analytical} to devise a single-layer perfect-electric-conductor (PEC)-backed reflecting MG for controlling several reflected modes using several meta-atoms \cite{popov2018controlling,popov2019constructing,popov2019designing}. Simultaneously, we have proposed the first conceptual multilayered multielement MGs for perfect anomalous refraction, with three (horizontally and vertically displaced) capacitively-loaded wires per period in free space \cite{epstein2018eucap}, adjusted and utilized later by Packo \textit{et al.} for obtaining similar functionalities for acoustic waves \cite{packo2019inverse}. Nevertheless, this is the first time, to the best of our knowledge, that a systematic rigorous formulation of a general synthesis scheme for engineering both amplitude and phase of an arbitrarily-large number of (reflected and transmitted) diffracted modes via multilayered PCB MGs is presented and verified experimentally. Importantly, the introduced MG configuration and design approach retain a simple semianalytically resolved capacitively-loaded wire geometry for all meta-atoms, allowing scalability and versatility for implementing diverse applications, without resorting to full-wave optimization. Thus, we expect this methodology to enable further demonstrations of large-period MGs for more advanced field manipulation functionalities in reflection or transmission, establishing them as appealing alternatives for periodic metasurfaces.

\section{Theory}
\label{sec:Theory}
\subsection{Formulation}
\label{subsec:Formulation}
We consider a 2D ($\partial/\partial x=0$) $\Lambda$-periodic configuration bounded within the region $z\in\left[0,t_{N-1}\right]$, excited by a transverse electric (TE) polarized ($E_{z}=E_{y}=H_{x}=0$) plane wave $E_{x}^{\mathrm{inc}}(y,z)=E_{\mathrm{in}}e^{-jky\sin\mathrm{\mathrm{\theta_{in}}}}e^{-jkz\cos\mathrm{\theta_{in}}}\hat{x}$, with angle of incidence of $\mathrm{\theta_{in}}$ and amplitude $E_{\mathrm{in}}$ (Fig. \ref{Fig:Config}). 
The MG under consideration is composed of $N-2$, generally different, dielectric layers, with given permittivities $\varepsilon_{n}$ and permeabilities $\mu_n$ for the $n$th layer \footnote{Although the formalism is general when using $k_n$ and $\eta_n$ as defined, we assume here the laminates are non magnetic, with permeabilities $\mu_n=\mu_0$.}; the $n=1$ and $n=N$ layers are assumed to contain air, stretching to $z\to +\infty$ and $z \to -\infty$, respectively. The wavenumber and the wave impedance for each layer are given, respectively, by $k_{n}=\omega\sqrt{\mu_{n}\varepsilon_{n}}$ and $\eta_{n}=\sqrt{\mu_{n}\varepsilon_{n}}$. The planes $z=t_{n-1}$ and $z=t_{n}$ define the boundaries of the $n$th layer, $n=2,3,...,N-1$, and we choose the coordinate system such that $t_{1}=0$.
Within a period, $K$ different loaded wires, with impedances per unit length $\tilde{Z}_k$, are distributed along the layer interfaces; similar to \cite{epstein2018eucap}, they are displaced in the $y$ and the $z$ directions relative to one another, such that the location of the $k$th wire is $\left(y,z\right)=\left(d_k,h_k\right)$, and we denote the current carried by it as $I_{k}$ [Fig. \ref{Fig:Config} (a) and (b)].

According to the FB theorem, scattering from such a structure will incur coupling to a discrete set of FB modes, whose transverse wavenumbers are given
by $k_{t,m}=k_{1}\sin\mathrm{\theta_{in}}+2\pi m/\Lambda$ for the $m$th mode. The longitudinal wavenumbers corresponding to the $m$th mode in the $n$th layer are given by ${\beta_{m,n}=\sqrt{k_{n}^{2}-k_{t,m}^{2}}}$, $\Im\{\beta_{m,n}\}\leq 0$. Given the period $\Lambda$ and the incident angle $\mathrm{\theta_{in}}$, the indices of the propagating modes in the $n$th layer $m_{n,\mathrm{prop}}$ must satisfy \cite{rabinovich2018analytical}

\begin{equation}
\label{eq:num_prop_modes}
\begin{aligned}
-\frac{\Lambda}{2\pi}(k_n+k_1\sin\mathrm{\theta_{in}})<m_{n,\mathrm{prop}}<\frac{\Lambda}{2\pi}(k_n-k_{1}\sin\mathrm{\theta_{in}})
\end{aligned}
\end{equation}
Substituting $k_n=k_1=k_N$ in \eqref{eq:num_prop_modes} quantifies the overall number of propagating modes in the outermost layers as $2M=2\left(M_{-}+M_{+}+1\right)$, accounting for both reflected and transmitted modes, with $M_- = \lfloor \frac{\Lambda}{2\pi}(k_n+k_1\sin\mathrm{\theta_{in}})\rfloor$ and $M_+ = \lfloor \frac{\Lambda}{2\pi}(k_n-k_{1}\sin\mathrm{\theta_{in}}) \rfloor$. 

Our goal is to devise a MG that would couple the incoming power in its entirety to the various propagating modes following a prescribed partition, setting the complex electric field amplitude of the $m$th transmitted and reflected mode, respectively, to the desirable values of $E_{m}^\mathrm{tran}=|E_{m}^\mathrm{tran}|e^{j\varphi_{m}^\mathrm{tran}}$ and $E_{m}^\mathrm{ref}=|E_{m}^\mathrm{ref}|e^{j\varphi_{m}^\mathrm{ref}}$, 
$m\in[-M_{-},M_{+}]$; global power conservation thus requires that

\begin{equation}
\label{eq:Tot_power_conservation}
\begin{aligned}
\sum_{m=-M_{-}}^{M_{+}}\frac{\beta_{m,1}}{\beta_{0,1}}\bigg(\bigg|\frac{E_{m}^{\mathrm{tran}}}{E_{in}}\bigg|^{2}+\bigg|\frac{E_{m}^{\mathrm{ref}}}{E_{in}}\bigg|^{2}\bigg)=1.
\end{aligned}
\end{equation}
To control the coupling to these $2M$ modes as prescribed, we propose using $K \geq 2M$ capacitively-loaded conducting strips with varying widths $W_k$ [Fig. \ref{Fig:Config}(b)], positioned, as stated above, at $(y,z) = \left(d_{k},h_{k}\right)$ for the $k$th wire ($k=1,2,...,K$), where the horizontal and vertical displacements $d_{k}\in\left[-\Lambda/2,\Lambda/2\right)$ and $h_{k}\in\left\{t_1,t_2,...,t_{N-1}\right\}$ are defined with respect to the reference wire at $(d_{1}, h_{1})=(0,0)$. As shall be detailed in Sections \ref{subsec:Coupling} and \ref{subsec:passivity}, this choice would provide the sufficient number of degrees of freedom to obtain the desired diffraction control with a passive lossless device. 

As in \cite{ra2017meta,epstein2017unveiling,rabinovich2018analytical,epstein2018eucap,ikonen2007modeling}, we analyze the problem by expressing the fields in space as a superposition of two sets: one corresponding to the
incident field scattered off the stratified media configuration in the absence of the loaded wires (external field); the other is produced by the current induced in the wires by the polarizing impinging wave. In contrast to this previous work, however, dealing with the multilayer PCB MG configuration requires accounting for multiple reflections at the boundaries between the various layers, similar to \cite{felsen1994radiation, Chew1990,osipov2017modern,epstein2010impact,epstein2013thesis}. Specifically, the electromagnetic fields will be resolved by enforcing
the continuity of the tangential components across the layer interfaces and the radiation condition, with the additional source condition applicable to the wire grid contribution alone, imposing a discontinuity in the tangential magnetic fields due to the induced electric line currents.

\subsection{External field contribution}
\label{subsec:external_field}
As stated, we begin by considering the interaction between the incident field and the dielectric stack in the absence of the MG wires. To evaluate this so called external field contribution, we follow the recursive formulation scheme described in \cite{Chew1990,epstein2013thesis}. For completeness, we briefly present here the main steps and results. Hence, the
electric field in the $n$th layer is written as a superposition of forward and backward waves
\begin{equation}
\label{eq:general_field_PW}
\begin{aligned}
E_{n}^\mathrm{ext}(y,z)=A_{0,n}^\mathrm{ext}e^{-jk_{t,0}y-j\beta_{0,n}z}+B_{0,n}^\mathrm{ext}e^{-jk_{t,0}y+j\beta_{0,n}z}
\end{aligned}
\end{equation}
where we note that the incident field wavevector coincides with the fundamental ($m=0$) FB mode of the general configuration, which explains the subscript notation. Clearly,
the coefficients $A_{0,n}^\mathrm{ext}$ and $B_{0,n}^\mathrm{ext}$, respectively, are the complex amplitude of these forward and backward plane waves, and the wavevector components $k_{t,0}$ and $\beta_{0,n}$ inherently satisfy the wave equation in the $n$th layer and follow Snell's law at the layer interfaces \cite{felsen1994radiation,Chew1990,osipov2017modern}.

Applying the continuity conditions at each interface $t_n$, and considering an incident plane wave with amplitude $A_{0,1}^\mathrm{ext}=E_\mathrm{in}$, it can be shown that the field in the $n$th layer can be formulated as \cite{epstein2010impact,epstein2013thesis}
\begin{equation}
\label{eq:recursive_field_left_ext}
\begin{aligned}
E_{n}^{\mathrm{ext}}(y,z)\!=\!E_{\mathrm{in}}\!\left(\prod_{p=1}^{n-1}T_{0,p}\!\!\right)\!e^{-jk_{t,0}y}\left(e^{-j\beta_{0,n}z}+R_{0,n}e^{j\beta_{0,n}z}\right)
\end{aligned}
\end{equation}
where $R_{0,n}$ and $T_{0,n}$ stand for the \emph{total} reflection and transmission coefficients of the $0$th mode with respect to the interface $t_n$ in the forward direction, and are given by \cite{epstein2010impact,epstein2013thesis}

\begin{equation}
\label{eq:total_Reflection_Transmission_coeff_left_ext}
\begin{aligned}
R_{0,n}&=\frac{B_{0,n}}{A_{0,n}}=\frac{\Gamma_{0,n}+R_{0,n+1}e^{2j\beta_{0,n+1}t_{n}}}{1+\Gamma_{0,n}R_{0,n+1}e^{2j\beta_{0,n+1}t_{n}}}e^{-2j\beta_{0,n}t_{n}} \\
T_{0,n}&=\frac{A_{0,n+1}}{A_{0,n}}=\frac{(1+\Gamma_{0,n})e^{j\beta_{0,n+1}t_{n}}}{1+\Gamma_{0,n}R_{0,n+1}e^{2j\beta_{0,n+1}t_{n}}}e^{-j\beta_{0,n}t_{n}}
\end{aligned}
\end{equation}
with the respective \emph{local} reflection coefficient in the forward direction defined as $\Gamma_{0,n}=\left(1-\gamma_{0,n}\right)/\left(1+\gamma_{0,n}\right)$, and the corresponding wave-impedance ratio as $\gamma_{0,n}=Z_{0,n}/Z_{0,n+1}$, with the TE wave impedance of the $m$th mode in the $n$th layer defined as $Z_{m,n}=k_n\eta_n/\beta_n$. Acknowledging that $\Gamma_{0,n}$ is the standard Fresnel coefficient for a plane wave impinging upon the interface $t_n$ with the wavevector $\vec{k}_{0,n}=k_{t,0}\hat{y}+\beta_{0,n}\hat{z}$, it can be shown that $R_{0,n}$ and $T_{0,n}$ properly account for all the multiple reflections (or, alternatively, the image series contribution) arising from the multiple interfaces $t_n,t_{n+1},...,t_{N-1}$ \cite{felsen1994radiation,Chew1990,osipov2017modern,epstein2010impact,epstein2013thesis}.
The stopping criterion for the recursion process is, naturally, $R_{0,N}=0$, manifesting the radiation condition at the $N$th region.

\subsection{Induced current contribution}
\label{subsec:Wire grid field}
Once the external field contribution is assessed via \eqref{eq:recursive_field_left_ext}-\eqref{eq:total_Reflection_Transmission_coeff_left_ext}, we proceed with a similar recursive formalism to compute the field contribution associated with the currents induced on the wire grid. Here we assume the $k$th wire at $(d_k,h_k)$ to carry (yet to be determined) induced current of $I_k$ parallel to the $x$ axis, forming an equivalent electric-line-source array. We evaluate the fields due to $I_k$ separately for every $k\in\left\{1,2,...,K\right\}$, and use superposition to obtain the response of the multilayered multielement wire grid. There are two fundamental differences between this problem and the one considered in Section \ref{subsec:external_field} that must be taken into consideration. First, the $I_k$ line-source array is $\Lambda$-periodic; thus, the fields would be scattered to an infinite number of FB modes. Second, the source condition should be formulated and enforced at $(d_k,h_k)$, introducing the expected discontinuity in the tangential magnetic fields. Once again, for brevity, we will not provide a detailed derivation herein; the results can be obtained via the modal analysis techniques presented in \cite{felsen1994radiation, Chew1990,osipov2017modern,epstein2010impact,epstein2013thesis} for a line source embedded in stratified media, with the discrete nature of the FB spectrum taken into account. For completeness, the key steps and essential modifications will be summarized below.

Correspondingly, the electric field in the $n$th layer due to an electric line source of current $I_k$ situated at $(d_k,h_k)$ can be expressed by a FB series
\begin{equation}
\label{eq:general_field_grid_FB}
\begin{aligned}
E_{n}^{(k)}(y,z)=\sum\limits_{m=-\infty}^{\infty}E_{m,n}^{(k)}(y,z)
\end{aligned}
\end{equation}
%
where the field associated with the $m$th mode can be written, by virtue of the wave equation, as a sum of forward and backward propagating modes
\begin{equation}
\label{eq:general_field_grid}
\begin{aligned}
E_{m,n}^{(k)}(y,z)&=A_{m,n}^{(k)}e^{-jk_{t,n}(y-d_{k})-j\beta_{m,n}z}\\
&+B_{m,n}^{(k)}e^{-jk_{t,n}(y-d_{k})+j\beta_{m,n}z}
\end{aligned}
\end{equation}
with respective amplitudes $A_{m,n}^{(k)}$ and $B_{m,n}^{(k)}$; the wavenumber components $k_{t,n}$ and $\beta_{m,n}$ are as defined in Section \ref{subsec:Formulation}.

We recall that the MG loaded wires, and consequently the induced currents, are situated at the interface $z=h_k=t_{n_k}$ between two media, the indices of which are $n_k$ and $\left(n_{k}+1\right)$ for the $k$th wire. Similar to \cite{Chew1990,epstein2010impact,epstein2013thesis}, we evaluate the fields using a recursive formulation, generalizing the one presented in Section \ref{subsec:external_field} for the fundamental mode. Due to the presence of the source, however, we use different recursion relations for the layers above ($n\geq n_k+1$) and below ($n\leq n_k$) the loaded wire, and utilize the source condition to tailor the solution at $z=h_k=t_{n_k}$.
%
Therefore, applying the continuity conditions for the
transverse components of the electric and magnetic fields at the interfaces $t_n$, $n>n_k$, and considering the orthogonality of the modes, we can resolve the modal fields for $z>h_k$ as
%
%
\begin{equation}
\label{eq:recursive_field_left}
\begin{aligned}
E_{m,n}^{(k)}(y,z)&=A_{m,n_k+1}^{(k)}\left(\prod_{p=n_k+1}^{n-1}T_{m,p}\right)\\
&\cdot e^{-jk_{t,m}(y-d_k)}\left(e^{-j\beta_{m,n}z}+R_{m,n}e^{j\beta_{m,n}z}\right)
\end{aligned}
\end{equation}
%
where the \emph{total} reflection and transmission coefficients of the $m$th mode with respect to the interface $t_n$ in the \emph{forward} direction are given by a generalization of \eqref{eq:total_Reflection_Transmission_coeff_left_ext}
\begin{equation}
\label{eq:total_Reflection_coeff_left}
\begin{aligned}
\!\!\!R_{m,n}&\!\!=\!\!\frac{B_{m,n}}{A_{m,n}}\!\!=\!\!\frac{\Gamma_{m,n}+R_{m,n+1}e^{2j\beta_{m,n+1}t_{n}}}{1+\Gamma_{m,n}R_{m,n+1}e^{2j\beta_{m,n+1}t_{n}}}e^{-2j\beta_{m,n}t_{n}} \\
\!\!\!T_{m,n}&\!\!=\!\!\frac{A_{m,n+1}}{A_{m,n}}\!\!=\!\!\frac{(1+\Gamma_{m,n})e^{j\beta_{m,n+1}t_{n}}}{1+\Gamma_{m,n}R_{m,n+1}e^{2j\beta_{m,n+1}t_{n}}}e^{-j\beta_{m,n}t_{n}}
\end{aligned}
\end{equation}
%
%
taking into account multiple reflections at the interfaces $t_n$, $n>n_k$, with the \emph{local} reflection coefficients for the $m$th mode in the \emph{forward} direction are given, once again, by the Fresnel formula $\Gamma_{m,n}=\left(1-\gamma_{m,n}\right)/\left(1+\gamma_{m,n}\right)$, where $\gamma_{m,n}=Z_{m,n}/Z_{m,n+1}$ and $Z_{m,n}=k_n\eta_n/\beta_n$ \cite{felsen1994radiation,Chew1990,osipov2017modern,epstein2010impact,epstein2013thesis}. The stopping criterion for this recursion formulation is $R_{m,N}=0$, as clearly no mode is reflected from the (non-existent) interface $t_N$.

Similarly, applying the continuity conditions at the interfaces $t_n$, $n<n_k$, yields the expressions for the fields at $z<h_k$, reading \cite{Chew1990,epstein2013thesis}
%
\begin{equation}
\label{eq:recursive_field_right}
\begin{aligned}
E_{m,n}^{(k)}(y,z)&=B_{m,n_k}^{(k)}\left(\prod_{p=n+1}^{n_k}\hat{T}_{m,p}\right)\\
&\cdot e^{-jk_{t,m}(y-d_k)}\left(e^{j\beta_{m,n}z}+\hat{R}_{m,n}e^{-j\beta_{m,n}z}\right)
\end{aligned}
\end{equation}
%
%
%
where the \emph{total} reflection and transmission coefficients of the $m$th mode with respect to the interface $t_{n-1}$ in the \emph{backward} direction are given by
\begin{equation}
\label{eq:total_Reflection_coeff_right}
\begin{aligned}
\hat{R}_{m,n}&\!\!=\!\!\frac{A_{m,n}}{B_{m,n}}\!\!=\!\!\frac{\hat{\Gamma}_{m,n}+\hat{R}_{m,n-1}e^{-2j\beta_{m,n-1}t_{n-1}}}{1+\hat{\Gamma}_{m,n}\hat{R}_{m,n-1}e^{-2j\beta_{m,n-1}t_{n-1}}}e^{2j\beta_{m,n}t_{n-1}} \\
\hat{T}_{m,n}&\!\!=\!\!\frac{B_{m,n-1}}{B_{m,n}}\!\!=\!\!\frac{(1+\hat{\Gamma}_{m,n})e^{-j\beta_{m,n-1}t_{n-1}}}{1+\hat{\Gamma}_{m,n}\hat{R}_{m,n-1}e^{-2j\beta_{m,n-1}t_{n-1}}}e^{j\beta_{m,n}t_{n-1}}
\end{aligned}
\end{equation}
%
%
and the \emph{local} reflection coefficients in the \emph{backward} direction are defined as $\hat{\Gamma}_{m,n}=\left(1-\hat{\gamma}_{m,n}\right)/\left(1+\hat{\gamma}_{m,n}\right)$, where for the backward case $\hat{\gamma}_{m,n}=Z_{m,n}/Z_{m,n-1}$, and the stopping criterion is $\hat{R}_{m,1}=0$. In analogy to the total reflection and transmission coefficients in the forward direction, the ones associated with the backward direction consider multiple reflections in the interfaces $t_{1}, t_2, ..., t_{n-1}$.
%

Equations \eqref{eq:recursive_field_left} and  \eqref{eq:recursive_field_right} demonstrate that to fully resolve the fields everywhere in space we required the knowledge of two coefficients, namely, the amplitudes of the forward propagating wave above the line source $A_{m,n_k+1}^{(k)}$ and the backward propagating wave below the line source $B_{m,n_k}^{(k)}$. To compute these, we enforce the remaining conditions at $z=h_k$: continuity of the tangential electric field, and the source condition for the tangential magnetic field due to the effective surface current. Specifically, by using the Poisson formula to transform the periodic line source array ("impulse train") into a modal sum, and using impulse balance techniques, the two conditions can be formulated for the $m$th mode as
\begin{equation}
\label{eq:source_condition_magnetic_field_wire}
\begin{aligned}
&\left.E_{m,n_k}^{(k)}(y,z)\right|_{z\rightarrow h_k^-}=\left.E_{m,n_k+1}^{(k)}(y,z)\right|_{z\rightarrow h_k^+}\\
&\left.H_{m,n_k}^{(k)}(y,z)\right|_{z\rightarrow h_k^-}-\left.H_{m,n_k+1}^{(k)}(y,z)\right|_{z\rightarrow h_k^+}=\frac{I_{k}}{\Lambda}
\end{aligned}
\end{equation}
where $H_{m,n}^{(k)}(y,z)=\frac{j}{k_n\eta_n}\frac{\partial}{\partial z} E_{m,n}^{(k)}(y,z)$, which form two linear equations for $A_{m,n_k+1}^{(k)}$ and $B_{m,n_k}^{(k)}$. Solving these with the aid of \eqref{eq:recursive_field_left} and \eqref{eq:recursive_field_right} yields explicit expressions for the two coefficients
\begin{equation}
\label{eq:source_condition_explicit_coefficients}
\begin{aligned}
&A_{m,n_k+1}^{(k)}=-\frac{I_k}{2\Lambda}Z_{m,n_k}\frac{1+\hat{R}_{m,n_k}e^{-2j\beta_{m,n_k}t_{n_k}}}{1-R_{m,n_k}\hat{R}_{m,n_k}}T_{m,n_k}\\
&B_{m,n_k}^{(k)}=-\frac{I_k}{2\Lambda}Z_{m,n_k+1}\frac{1+{R}_{m,n_k+1}e^{2j\beta_{m,n_k+1}t_{n_k}}}{1-\hat{R}_{m,n_k+1}{R}_{m,n_k+1}}\hat{T}_{m,n_k+1}
\end{aligned}
\end{equation}
which allow complete evaluation of the fields produced by the current $I_k$ induced on the $k$th wire, using \eqref{eq:general_field_grid_FB}, \eqref{eq:recursive_field_left}, and  \eqref{eq:recursive_field_right}.

Finally, for given stratified media configuration, incident plane wave, and the currents on the various loaded wires $I_k$, $k=1,2,...,K$, the total field in the $n$th layer can be deduced by superposition to be [\eqref{eq:recursive_field_left_ext}, \eqref{eq:general_field_grid_FB}]
\begin{equation}
\label{eq:Total_electric_field_space}
\begin{aligned}
E_{n}^{\mathrm{tot}}(y,z)=E_{n}^{\mathrm{ext}}(y,z)+\sum\limits_{k=1}^{K}\sum\limits_{m=-\infty}^{\infty}E_{m,n}^{(k)}(y,z)
\end{aligned}
\end{equation}

\subsection{Coupling to diffracted modes}
\label{subsec:Coupling}
The total field in the observation regions, given by \eqref{eq:Total_electric_field_space} with $n=1$ and $n=N$, depends on the current $I_k$ induced on the various wires 
and on their positions $\left(d_k,h_k\right)$, as manifested, e.g., by \eqref{eq:recursive_field_left}, \eqref{eq:recursive_field_right}, and \eqref{eq:source_condition_explicit_coefficients}. 
Therefore, in order to achieve the desired coupling to the various reflected and transmitted modes, we use the analytical model to determine the individual currents required to yield the prescribed modal field amplitudes and phases as defined in \eqref{eq:Tot_power_conservation}; later on, we will indicate how to guarantee that these optimal currents would indeed be induced on the MG passive conductors by proper design of the distributed loads and wire relative offsets.

More specifically, the requirement for particular transmitted and reflected modal amplitudes $E_{m}^\mathrm{tran}=|E_{m}^\mathrm{tran}|e^{j\varphi_{m}^\mathrm{tran}}$ and $E_{m}^\mathrm{ref}=|E_{m}^\mathrm{ref}|e^{j\varphi_{m}^\mathrm{ref}}$ forms a set of linear constraints tying the various MG degrees of freedom, similar to the specular reflection elimination condition derived in \cite{ra2017meta,epstein2017unveiling,rabinovich2018analytical,epstein2018eucap}. Thus, we can formulate a set of $2M$ linear modal equations expressing the desired total field for each mode: $M$ equations to control the reflected modes, reading 
\begin{equation}
\label{eq:linear_eq_reflection}
\begin{aligned}
&B_{0,1}^\mathrm{ext}\delta_{m,0} + \sum\limits_{k=1}^{K} B_{m,1}^{(k)}e^{jk_{t,1}d_k} = E_m^\mathrm{ref}
\end{aligned}
\end{equation}
and $M$ equations to control the transmitted modes, namely,
\begin{equation}
\label{eq:linear_eq_transmission}
\begin{aligned}
&A_{0,N}^\mathrm{ext}\delta_{m,0} + \sum\limits_{k=1}^{K} A_{m,N}^{(k)}e^{jk_{t,N}d_k} = E_m^\mathrm{trans}\\
\end{aligned}
\end{equation}
where the propagating mode indices are ${m=-M_-, ..., 0, ..., M_+}$, and the coefficients $A_{m,N}$ and $B_{m,1}$ depend on the wire positions and induced currents as formulated in Sections \ref{subsec:external_field} and \ref{subsec:Wire grid field}.

%
Equations \eqref{eq:general_field_grid}, \eqref{eq:recursive_field_left}, \eqref{eq:recursive_field_right}, and \eqref{eq:source_condition_explicit_coefficients} indicate that each of the coefficients $A_{m,N}^{(k)}$ and $B_{m,1}^{(k)}$ is proportional to the current on the $k$th wire. Expressing these coefficients of proportionality as impedances per unit length results with the next matrix equation:
\begin{equation}
\label{eq:Matrix_equation}
\begin{aligned}
\begin{pmatrix}
  \mathbf{Z^\mathrm{ref}_{M\times K}}   \vspace{3pt} \\
  \mathbf{Z^\mathrm{trans}_{M\times K}}
  \end{pmatrix}
  \begin{pmatrix}
  \mathbf{I_{K\times 1}}
  \end{pmatrix}
  =  
  \begin{pmatrix}
  \mathbf{V^\mathrm{ref}_{M\times 1}}   \vspace{3pt} \\
  \mathbf{V^\mathrm{trans}_{M\times 1}}
  \end{pmatrix}
\end{aligned}
\end{equation}
where the reflection and transmission impedance matrix elements $Z^\mathrm{ref}_{m,k}$ and $Z^\mathrm{trans}_{m,k}$ relate the $m$th mode reflected or transmitted field amplitude to the current induced on the $k$th wire; the current vector $\mathbf{I_{K\times 1}}$ elements are these induced currents $I_k$; and the excitation vectors $\mathbf{V^\mathrm{ref}_{M\times 1}}$ and $\mathbf{V^\mathrm{trans}_{M\times 1}}$ are defined, respectively, as $V^\mathrm{ref}_m=E_{m}^\mathrm{ref}-\delta_{m,0}B_{0,1}^{\mathrm{ext}}$ and $V^\mathrm{trans}_m=E_{m}^\mathrm{trans}-\delta_{m,0}A_{0,N}^{\mathrm{ext}}$.
Finally, to obtain the distribution of induced currents establishing the desired modal field amplitudes, we can solve this equation by a regular matrix inversion in case we choose the number of wires per period to be $K=2M$, or by using a Moore-Penrose pseudoinverse method in case $K>2M$ is chosen.






Next, we need to assess the proper wire loading $\tilde{Z}_{k}$ to facilitate the desirable current induction, prescribed by \eqref{eq:Matrix_equation}. Similar to \cite{tretyakov2003analytical,epstein2017unveiling,rabinovich2018analytical,ikonen2007modeling}, this is achieved by using Ohm's law to relate the total fields applied on a wire to the current flowing through it via the distributed load impedance. Explicitly, for the $k$th wire situated at the interface $z=h_k=t_{n_k}$, Ohm's law reads
\begin{equation}
\label{eq:load_impedances}
\begin{aligned}
\tilde{Z}_{k}I_{k}&=E_{n_k}^{\mathrm{ext}}(d_{k},h_{k})+ \sum_{q\neq k} \sum\limits_{m=-\infty}^{\infty}E^{(q)}_{m,n_k}(d_{k},h_{k})\\
&+\sum_{m=-\infty}^{\infty}E^{(k)}_{m,n_k}(y\rightarrow d_k,z\rightarrow h_k)
\end{aligned}
\end{equation}
where the total field acting on the $k$th wire, written on the right hand side (RHS), features three main contributions. The first is the result of the excitation field interacting with the stratified media configuration \emph{in the absence} of the conducting wires (Section \ref{subsec:external_field}), and can be calculated via \eqref{eq:general_field_PW}-\eqref{eq:total_Reflection_Transmission_coeff_left_ext}. The second term corresponds to the field produced by all the current-carrying wires \emph{other than} the $k$th wire at the reference wire position $(y,z) = (d_{k},h_{k})$, and can be evaluated using \eqref{eq:general_field_grid}-\eqref{eq:source_condition_explicit_coefficients} (Section \ref{subsec:Wire grid field}). The third term accounts for the fields produced by the $k$th wire on its shell. Similar to \cite{tretyakov2003analytical,epstein2017unveiling,rabinovich2018analytical,ikonen2007modeling}, due to the singularity of the Hankel function at the origin, it is not possible to use \eqref{eq:general_field_grid}-\eqref{eq:source_condition_explicit_coefficients} as is at $\left(y,z\right)\rightarrow\left(d_k,h_k\right)$, and a more subtle treatment is required. 



To facilitate resolution of these terms as well, we follow the technique detailed in our previous report \cite{rabinovich2018analytical}. In particular, we \emph{separate} the term corresponding to the modal decomposition of the line source formed by the current-carrying $k$th wire $E^{(k)}_{\mathrm{self,direct}}$ from the multiple reflection series $\sum_{m=-\infty}^{\infty}E^{(k)}_{m,n_k}$, and evaluate it separately from the rest of the sum, which we denote as the image series contribution $E^{(k)}_{\mathrm{self,image}}$ \cite{felsen1994radiation,Chew1990}. 
As the singularity only appears in $E^{(k)}_{\mathrm{self,direct}}$, we evaluate it on the wire shell and utilizing the flat wire approximation \cite{tretyakov2003analytical}
\begin{equation}
\label{eq:Self_term}
\begin{aligned}
&E^{(k)}_{\mathrm{self,direct}}=\frac{k_{n_k}\eta_{n_k}I_{k}}{2}\Big\{\frac{1}{k_{n_k}\Lambda \cos \theta_\mathrm{in}}+\frac{j}{\pi}\Big[\log \frac{2\Lambda}{\pi w}\\
&+\frac{1}{2}\sum_{\substack{m=-\infty \\ m\neq 0}}^{\infty}\Big(\frac{2\pi}{\sqrt{(2\pi m+k_{1}\Lambda\sin \theta_\mathrm{in})^{2}-(k_{n_k}\Lambda)^{2}}}-\frac{1}{\left| m \right|}\Big)\Big]\Big\}
\end{aligned}
\end{equation}
%
%
$w$ being the conductor width [Fig. \ref{Fig:Config}(c)], while for the image series field we can safely substitute $\left(y,z\right)=\left(d_k,h_k\right)$, yielding
\begin{equation}
\label{eq:image_field}
\begin{aligned}
E^{(k)}_{\mathrm{self,image}}&=\sum_{m}\bigg\{\frac{\eta_{n_k}k_{n_k}}{2\Lambda\beta_{m,n_k}}I_{k}+\\
&+A_{m,n_k}^{(k)}e^{-j\beta_{m,n_k}h_k}+B_{m,n_k}^{(k)}e^{+j\beta_{m,n_k}h_k}\bigg\}
\end{aligned}
\end{equation}
such that overall, $\sum_{m=-\infty}^{\infty}E^{(k)}_{m,n_k}(y\rightarrow d_k,z\rightarrow h_k)=E^{(k)}_{\mathrm{self,direct}}+E^{(k)}_{\mathrm{self,image}}$ in \eqref{eq:load_impedances}.

Lastly, after evaluating the load impedances $\tilde{Z}_k$ of \eqref{eq:load_impedances} that would lead to the desired induced currents $I_k$ of \eqref{eq:Matrix_equation}, we harness our previous work to calculate the printed capacitor width $W_{k}$ [Fig. \ref{Fig:Config}(c)] that would realize the required distributed load impedance $\tilde{Z}_{k} =
1/ (j\omega lC_{k})$, where $C_{k}$ is the load capacitance of the $k$th wire, and $l$ is the periodicity along the $x$ axis \cite{epstein2017unveiling,rabinovich2018analytical}.
Specifically, when the trace width is equal to the gap between traces $w = s$ as in our case, we can approximate the capacitor width following $W_{k}=2.85 K_{\mathrm{corr}} C_{k}/\varepsilon_{\mathrm{eff,r}} \mathrm{[mil/fF]}$ \cite{guptamicrostrip}, where $K_{\mathrm{corr}}$ is a frequency dependent correction factor (at $20\mathrm{GHz}$ and $w=s=4\mathrm{mil}$ utilized in Section \ref{sec:results} it was previously found that $K_\mathrm{corr}=0.947$), and the effective permittivity is given by $\varepsilon_\mathrm{eff,r}=(\varepsilon_{n_k}+\varepsilon_{n_k+1})/(2\varepsilon_0)$, where $z=h_k=t_{n_k}$ is the vertical coordinate of the wire \cite{rabinovich2018analytical}.

\subsection{Passivity condition and loss minimization}
\label{subsec:passivity}
Apparently, the outlined procedure facilitates analytical design of the MG in mind, exercising full control over the $2M$ (forward and backward) propagating FB modes in the considered configuration. For given stratified media $\varepsilon_n$, $t_n$, meta-atom distribution ($d_{k}, h_{k}$), and desirable modal coefficients $E_{m}^\mathrm{ref}$ and $E_{m}^\mathrm{tran}$, we retrieve the required wire currents via \eqref{eq:Matrix_equation}, subsequently calculate the distributed load impedances $\tilde{Z}_k$ to induce them via \eqref{eq:load_impedances}, from which the capacitor geometries $W_{k}$ can be found; these yield the complete trace specifications comprising the MG
design. Nonetheless, this scheme implicitly
relies on two critical assumptions, the validity of which depends on the judicious selection of the MG layout ($d_{k}$,$h_{k}$). First, it is assumed that the derived load impedances would be completely reactive, such that they could be realized via passive and lossless printed capacitors.
While such a solution is principally feasible, other designs, relying on balanced loss and gain in different meta-toms,
could also facilitate the desirable response, thus have to be rejected explicitly \cite{epstein2018eucap}. Second, our synthesis method further assumes that conductor losses, which are neglected during the formulation but are inevitable in a realistic implementation, would not deteriorate much the MG performance. However, this parasitic power dissipation depends greatly on the magnitude of
currents developing on the various wires \cite{epstein2017unveiling,rabinovich2018analytical}, which should, thus, be minimized to retain the model fidelity.

These two requirements of the MG are translated into two sets of \emph{nonlinear} physical constraints, enforced by suitable tuning of the meta-atom positions
($d_{k}$,$h_{k}$), which form additional $2(K-1)$ degrees of freedom that have not been exploited yet. As per the discussion in the previous paragraph, the first set of conditions should guarantee the MG would not require lossy and active meta-atoms to satisfy \eqref{eq:Matrix_equation} with the prescribed (complex) modal field amplitudes
$E_{m}^\mathrm{ref}$ and $E_{m}^\mathrm{tran}$. Although it is tempting to require the minimization
of $|\Re\{Z_{k}\}|/|\Im\{Z_{k}\}|$ for all meta-atoms, thus ensuring a dominant reactive response, this condition will not necessarily lead to the desirable outcome. The problem stems from the fact that using this condition may lead to solutions in which both $|\Re\{Z_{k}\}|$ and $|\Im\{Z_{k}\}|$ are large, thus placing no limit on the real power exchanged with the loaded wires by design (recall that we wish the meta-atoms to (ideally) be completely passive and lossless). Instead, therefore, we formulate a constraint on the power, rather than the impedance, demanding that the power
dissipated or introduced by each of the meta-atoms \emph{by design} $P_{k}^\mathrm{dsn}$ would be negligible with respect to the incident
power $P_{\mathrm{in}}$. This would ensure that the electromagnetic field solution expects little power to be provided or absorbed by the meta-atoms to sustain the desirable field
pattern, and thus would still be valid even if the real part of the impedance specified by the synthesis procedure will be neglected. Formally, this passivity condition corresponds to the following set of $K-1$ constraints

\begin{equation}
\label{eq:passivity_condition}
\begin{aligned}
a_{k}\!=\!\frac{P_{k}^{\mathrm{dsn}}}{P_{\mathrm{in}}}\!=\!\frac{\frac{1}{2}\frac{|\Re\{Z_{k}\}||I_{k}|^{2}}{\Lambda}}{\frac{1}{2}\frac{\beta_{0}}{k_{1}}\frac{|E_{in}|^{2}}{\eta_{1}}}\!\rightarrow\!0, \,\,\,\,\,\,k\!=\!2,3,\dots,{K}
\end{aligned}
\end{equation}
The passivity of the meta-atom in the origin $k=1$ is guaranteed by global power conservation \eqref{eq:Tot_power_conservation}.

The second set of constraints is intended to retain the MG functionality in the presence of realistic conductor losses. Once more, as these conductor losses are neglected as part of the design procedure [Eq. \eqref{eq:Tot_power_conservation} implies that the incident power is distributed among the various propagating FB modes without any dissipation], our goal is to reduce the fraction of power absorbed in the MG in practice as much as possible. Following the same rationale leading to \eqref{eq:passivity_condition}, if the power balance envisioned by the analytical model will not be affected considerably by this unintentional dissipation, it is expected that the predicted field pattern will be formed as planned. Of course, this would also contribute to enhanced power efficiency. Therefore, the other set of constraints, minimizing the MG absorption, can be formulated as
\begin{equation}
\label{eq:lossless_condition}
\begin{aligned}
b_{k}\!=\!\frac{P_{k}^{\mathrm{loss}}}{P_{\mathrm{in}}}\!=\!\frac{\frac{1}{2}\frac{|I_{k}|^{2}\delta \tilde{R}}{\Lambda}}{\frac{1}{2}\frac{\beta_{0}}{k_{1}}\frac{|E_{in}|^{2}}{\eta_{1}}}\rightarrow0,\,\,\,\, k=1,2,3,\dots,K
\end{aligned}
\end{equation}
where $\delta \tilde{R}$ is the distributed resistance of the printed capacitors, which can be estimated based on the conductor width, conductivity, and the skin depth at the operation frequency as per \cite{epstein2017unveiling}, mostly independent of the capacitor width $W_{k}$. In fact, we transform these into $K-1$ constraints by averaging every two consecutive parameters, thus minimizing $(b_{k} + b_{k-1})/2$, for $k = 2,3,\dots K$. We emphasize that this step is critical for obtaining functional devices with good performance. The ability to formulate the corresponding conditions analytically, made possible with the loss analysis introduced in \cite{epstein2017unveiling}, further highlights the strength and merit of the model utilized herein.

Finally, we can lay out the complete synthesis procedure.
To design the envisioned MG, we solve the set of $K$ linear equations in \eqref{eq:Matrix_equation} under the $2(K-1)$ nonlinear constraints defined by \eqref{eq:passivity_condition} and \eqref{eq:lossless_condition}. This is implemented using the library function \texttt{lsqnonlin} in MATLAB, looking for the optimal wire positions ($d_{k}$,$h_{k}$) that yield induced currents $I_{k}$ [Eq. \eqref{eq:Matrix_equation}] and load impedances $\tilde{Z}_{k}$ [Eq. \eqref{eq:load_impedances}] that best satisfy the passivity condition [Eq. \eqref{eq:passivity_condition}] while minimizing power dissipation in a realistic MG [Eq. \eqref{eq:lossless_condition}]. These $3K-2$ constraints correspond to the $3K-2$ degrees of freedom in our configuration, featuring $K$ meta-atoms: the individual load impedances, and the horizontal and vertical spatial offsets with respect to the reference wire at the origin. Once the optimal $(d_{k}$,$h_{k})$ are determined, they are substituted into \eqref{eq:Matrix_equation} and \eqref{eq:load_impedances} to retrieve the load impedances (and subsequently the capacitor widths) that would facilitate the prescribed diffraction pattern. This yields the final layout of the MG, namely, the position and dimensions of all conductor traces, without requiring even a single optimization in a full-wave solver.

\section{Results and discussion}
\label{sec:results}

\subsection{Perfect anomalous refraction}
\label{subsec:refractor}
\begin{figure}[t]
\centering
\includegraphics[width=2.5in]{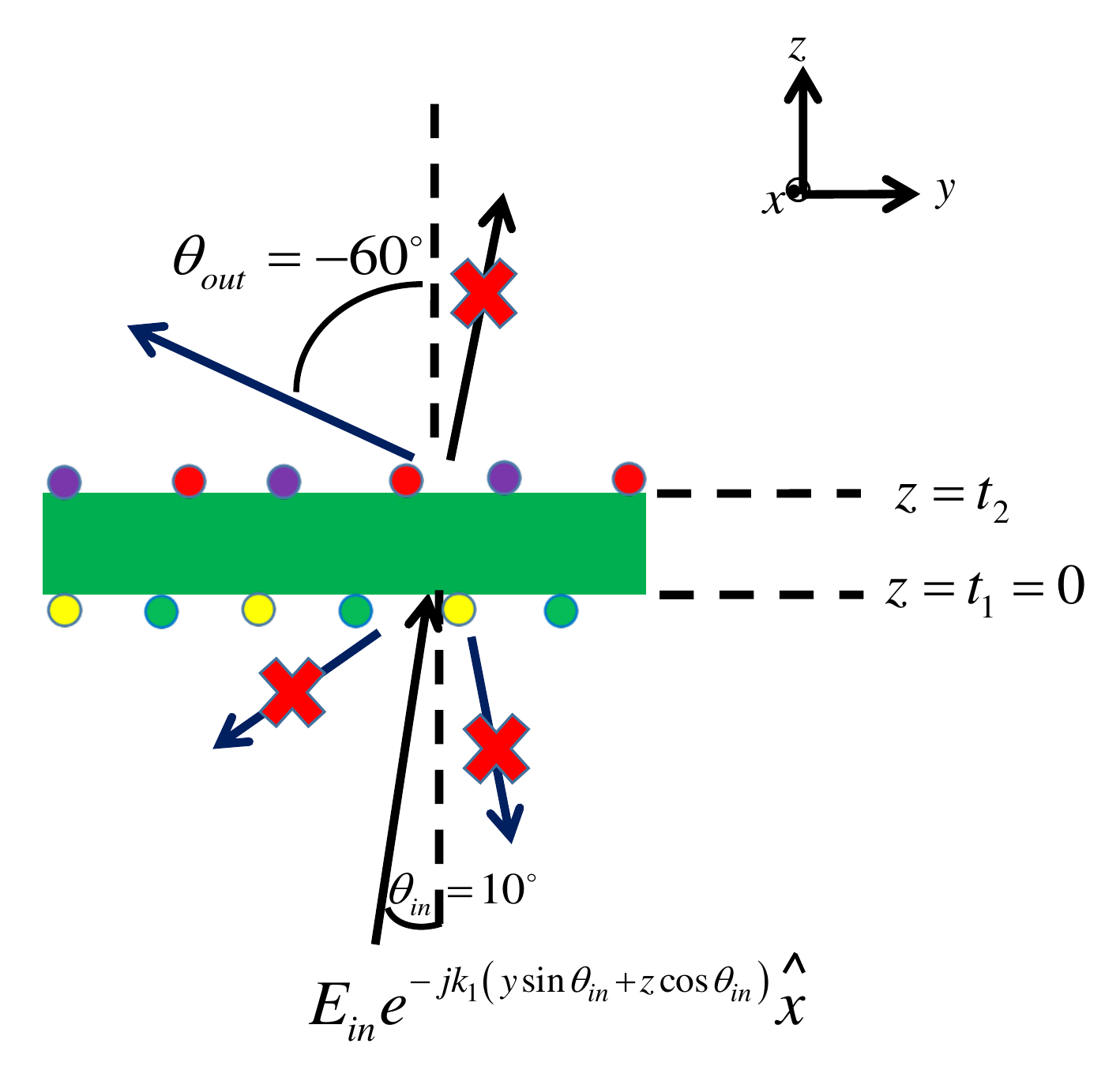}
\caption{Physical configuration of the MG prototype implementing perfect anomalous refraction, consisting of 4 meta-atoms arranged in 2 layers.}
\label{Fig:Config_Refractor}
\end{figure}

To verify our methodology, we follow the theoretical scheme described in Section \ref{sec:Theory} to first design a MG implementing anomalous refraction towards a prescribed angle. This example is of importance, as it marks, to the best of our knowledge, the first holistic demonstration, from semianalytical theory to experimental validation, of a practical multilayered multielement MG for manipulating diffraction modes in transmission. 
Specifically, we consider a plane wave at $f=20\,\mathrm{GHz}$ impinging from $\mathrm{\theta_{in}}=10^{\circ}$ and our aim is to couple the incident power in its entirety towards an angle of $\mathrm{\theta_{out}}=-60^{\circ}$ (Fig. \ref{Fig:Config_Refractor}). The difference between the transverse wave momenta dictates the periodicity along $y$, yielding $\Lambda=\lambda/\left|\sin\theta_\mathrm{in}-\sin\theta_\mathrm{out}\right|=0.96 \lambda=14.418\,\mathrm{mm}$. In this case, only two modes in transmission and two modes in reflection are propagationg as per \eqref{eq:num_prop_modes}. Thus, we need at least $4$ elements per period to establish the required FB mode manipulation. 

Subsequently, we propose using a MG configuration composed of a single dielectric slab with two meta-atoms on each of its facets, corresponding to $N=3$ and $K=4$ in Fig. \ref{Fig:Config}(a). The chosen substrate is Rogers RO3003 ($\varepsilon_{2}=3\varepsilon_0$ and $\tan\delta=0.001$) and we fix its thickness to the commercially available value of $H=60\,\mathrm{mil}=1.52\,\mathrm{mm}\approx\lambda/10$. 
This choice reveals a principal issue, encountered when the described MG design procedure faces the constraints over standard laminate thickness and permittivity as offered by commercial suppliers. In contrast to scenarios such as \cite{epstein2018eucap}, these practical constraints effectively reduce the available degrees of freedom, as they limit the vertical coordinates possible for the meta-atoms $h_k$ (limited to interfaces between laminates, top, or bottom facets). 

There are several ways to cope with this limitation; in this paper, we chose to relax our demands with respect to the passivity condition and loss minimization [\eqref{eq:passivity_condition} and \eqref{eq:lossless_condition}], such that we form a compacted set of nonlinear constraints, matching the reduced number of degrees of freedom, hence resolvable using \texttt{lsqnonlin}. More specifically, we attempt minimizing the \emph{mean} passivity deviation or absorption of several meta-atoms together, e.g., minimizing $(b_1+b_2+b_3+b_4)/4$ as a single constraint instead of minimzing $b_1$,$b_2$,$b_3$ and $b_4$ as four independent constraints (see Appendix \ref{app:optimization} for full implementation details for each of the considered prototypes). 
At the same time, the overall design goal is not affected, as all $a_k$ and $b_k$ are non-negative quantities [see \eqref{eq:passivity_condition} and \eqref{eq:lossless_condition}]. 
For the anomalous refraction case under consideration, the degrees of freedom relevant for these nonlinear constraints are the horizontal offsets of the meta-atoms, $d_2$, $d_3$, $d_4$; thus, three constraints are formed from \eqref{eq:passivity_condition} and \eqref{eq:lossless_condition} (see Appendix \ref{app:optimization}).

Once we have set the configuration and identified the associated degrees of freedom, we "encode" the desired functionality into the design mechanism by suitable definition of the required modal field amplitudes \eqref{eq:Tot_power_conservation}. To achieve perfect coupling to the (anomalous refraction) $m=-1$ FB mode in transmission, we set $E_{0}^\mathrm{ref}=E_{-1}^\mathrm{ref}=E_{0}^\mathrm{trans}=0$ and $E_{-1}^\mathrm{trans}=E_\mathrm{in}\sqrt{\beta_{0,1}/\beta_{-1,1}}$, and substitute them into the RHS of \eqref{eq:Matrix_equation}.
%
%
For given meta-atom coordinates $(d_k,h_k)$ and laminate permittivity, the impedance matrices on the LHS can be evaluated, and the induced currents $I_k$ can be resolved by matrix inversion. These, in turn, enable assessment of the necessary load impedances $\tilde{Z}_k$ via \eqref{eq:load_impedances}, facilitating calculation of the deviation from passivity and expected loss $a_k$ and $b_k$ [\eqref{eq:passivity_condition} and \eqref{eq:lossless_condition}]. This formulation of the forward problem, namely, tying a given wire distribution with the parameters we wish to minimize, paves the path for harnessing MATLAB to find the preferable values for the available degrees of freedoms (the horizontal offsets of the meta-atoms with respect to each other $d_2$, $d_3$, and $d_4$), yielding nearly-passive design requirements with minimal expected absorption. Once the optimal positions are obtained, the same forward problem formalism can be used to derive the required load impedance (and subsequently capacitor width) that would implement the designated functionality.
The results of this MATLAB optimization for the device considered herein, prescribing the detailed fabrication-ready PCB layout for the anomalous refraction MG, are summarized in Table \ref{tab:Refractor}. 

\begin{table}[t]
\centering
\begin{threeparttable}[b]
\renewcommand{\arraystretch}{1.3}
\caption{Design specifications of the MG for perfect anomalous refraction (corresponding to Fig. \ref{Fig:Config_Refractor}).} 

\label{tab:Refractor}
\centering
\begin{tabular}{|c||c|c||c|c|}
\hline \hline

& \multicolumn{2}{c||}{1st layer}  &  \multicolumn{2}{c|}{2nd layer} \\	\hline
& $k=1$ & $k=2$ & $k=3$ & $k=4$ \\ \hline \hline
$h_k [\lambda]$ & $0$ & $0$ & $0.101$ & $0.101$  \\	\hline
$d_k [\lambda]$ & $0$ & $-0.469$ & $-0.011$ & $0.438$  \\	\hline
$W_k [\mathrm{mm}]$ & $1.4162$ & $1.5483$ & $1.7498$ & $2.111$  \\	\hline 
\hline 
\end{tabular}
\end{threeparttable}
\end{table}

\begin{figure}[t]
\centering
\includegraphics[width=2.1in]{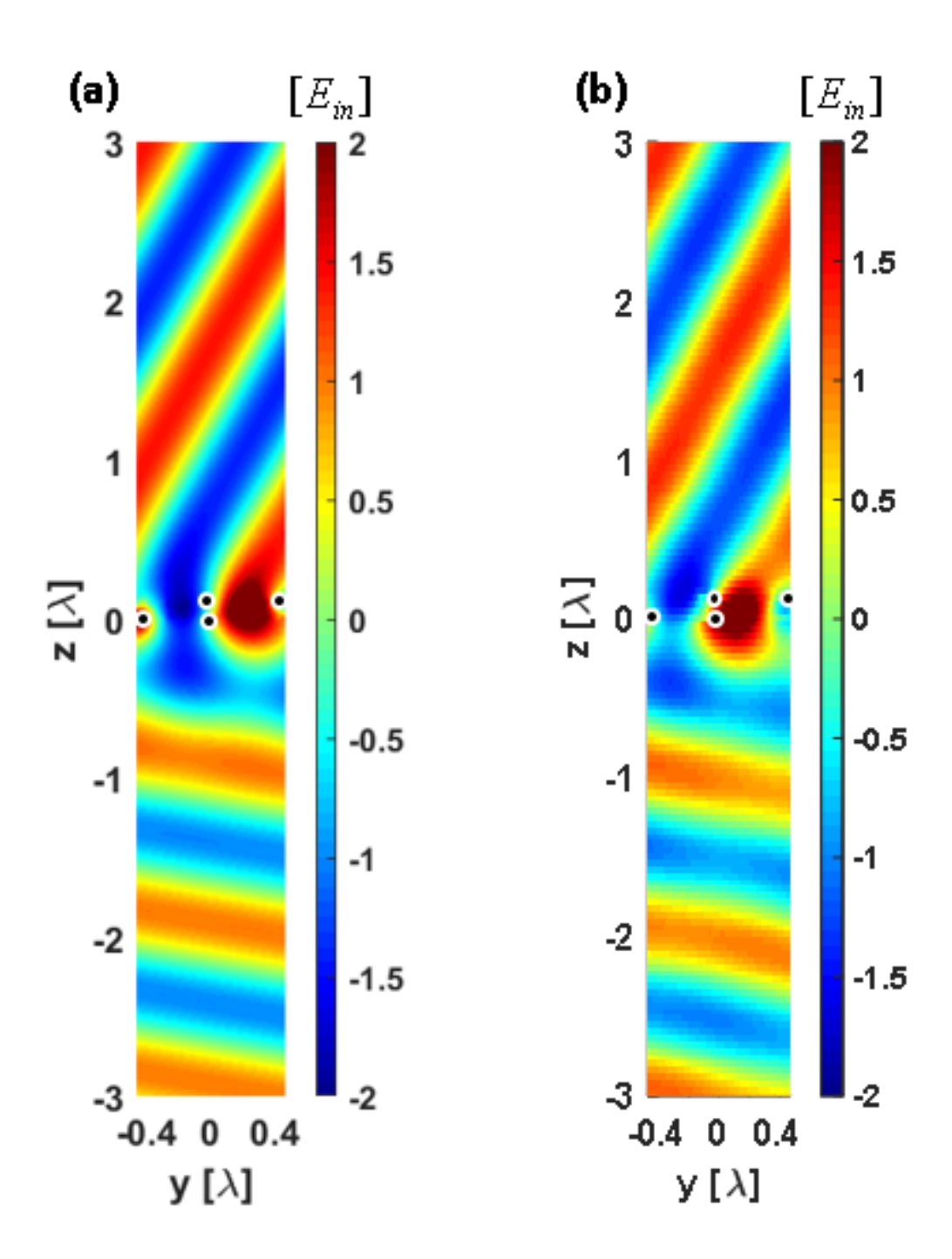}
\caption{Electric field distributions $\Re\{E_{x}(y,z)\}$ corresponding to the MG prototype implementing perfect anomalous refraction when excited from $\mathrm{\theta_{in}}=10^{\circ}$ at $f=20 \mathrm{GHz}$, (a) as predicted by the analytical model and (b) as recorded by full-wave simulations. 
A single period of the MG is presented; white circles denote the positions of the meta-atoms.}
\label{Fig:Field_Refractor}
\end{figure}

The printed conductors and dielectric substrate were correspondingly defined in CST Microwave Studio and simulated under periodic boundary conditions. Figure \ref{Fig:Field_Refractor} compares between the electric field distribution $\Re \{E_{x}(y,z)\}$ as obtained from the analytical model [Fig. \ref{Fig:Field_Refractor}(a)]
and as recorded in full-wave simulation [Fig. \ref{Fig:Field_Refractor}(b)], indicating excellent agreement between the two (white circles mark the locations of the $4$ meta-atoms). Full-wave simulations indicate that $89\%$ of the incident power is coupled to the desirable mode propagating towards $\mathrm{\theta_{out}}=-60^{\circ}$, with $10.8\%$ absorbed due to inevitable conductor and dielectric losses, and as little as $0.2\%$ coupled to spurious modes, as designed. 

\begin{figure}[t]
\centering
\includegraphics[width=3.4in]{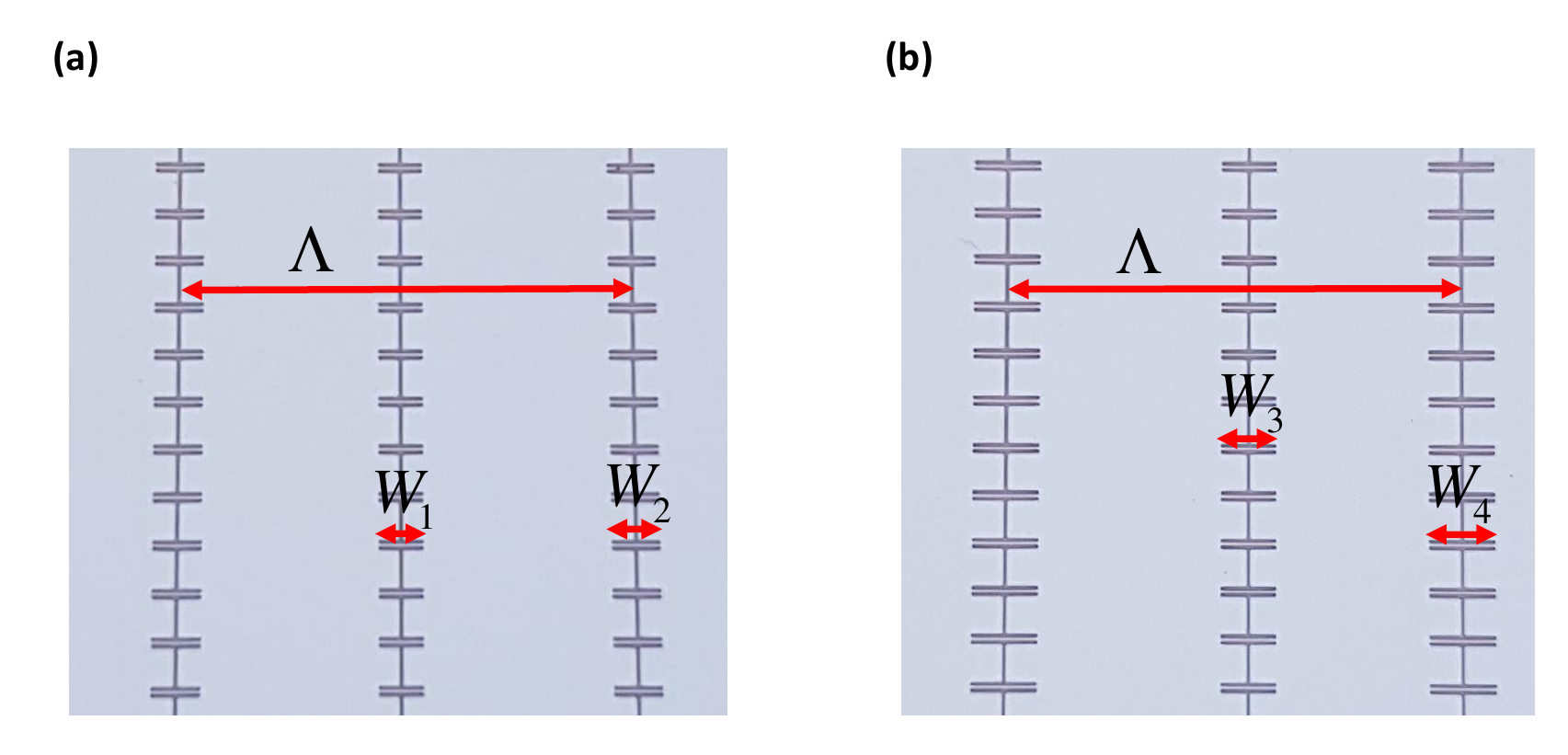}
\caption{Zoom-in on one lateral and twelve longitudinal periods of the fabricated MG prototype implementing perfect anomalous refraction. (a) top view; (b) bottom view. Capacitor widths correspond to Table \ref{tab:Refractor}.}
\label{Fig:Zoom_in_board}
\end{figure}

\begin{figure}[t]
\centering
\includegraphics[width=3.4in]{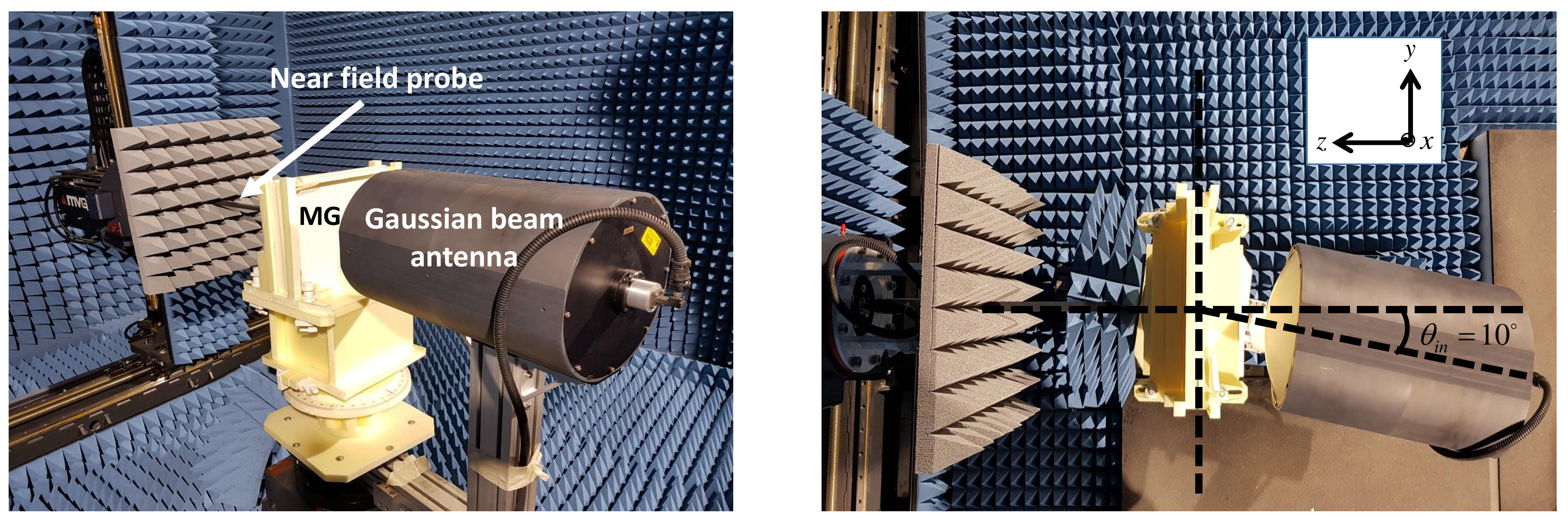}
\caption{Experimental setup used for characterizing the fabricated MGs, featuring a near field probe, a Gaussian beam antenna and the MG board, properly positioned in an anechoic chamber at the Technion.}
\label{Fig:Exp_setup}
\end{figure}

After this successful validation of the theoretical design in the full-wave solver, we turned to experimental verification of the MG performance. Conveniently, without any further optimization, we used the parameters listed in Table \ref{tab:Refractor} to fabricate a $9''\times12''$ board (PCB Technologies Ltd.), see Fig. \ref{Fig:Zoom_in_board}, and characterized in an anechoic chamber at the Technion. Within the chamber, the MG was mounted on a foam holder, placed at the focus of a Gaussian beam antenna (Millitech, Inc., GOA-42-S000094, focal distance of $196 \mathrm{mm}\approx 13\lambda$), illuminating the device under test (DUT) with a quasi-planar wavefront from the designated angle of incidence $\mathrm{\theta_{in}}=10^{\circ}$ (Fig. \ref{Fig:Exp_setup}). A planar near-field measurement system (MVG/Orbit-FR) was used to record the forward scattering pattern, scanning an area of 1.4m (in the $y$ direction) over 0.28m (in the $x$ direction) at a distance of $z=16.5\,\mathrm{cm}\approx11\lambda$ from the MG plane, from which the far field can be deduced using the equivalence principle \cite{balanis2012advanced}. For reference, we removed the MG board and measured the direct illumination of the Gaussian beam antenna on the same plane, allowing quantification of the overall excitation power.

\begin{figure}[t]
\centering
\includegraphics[width=2.5in]{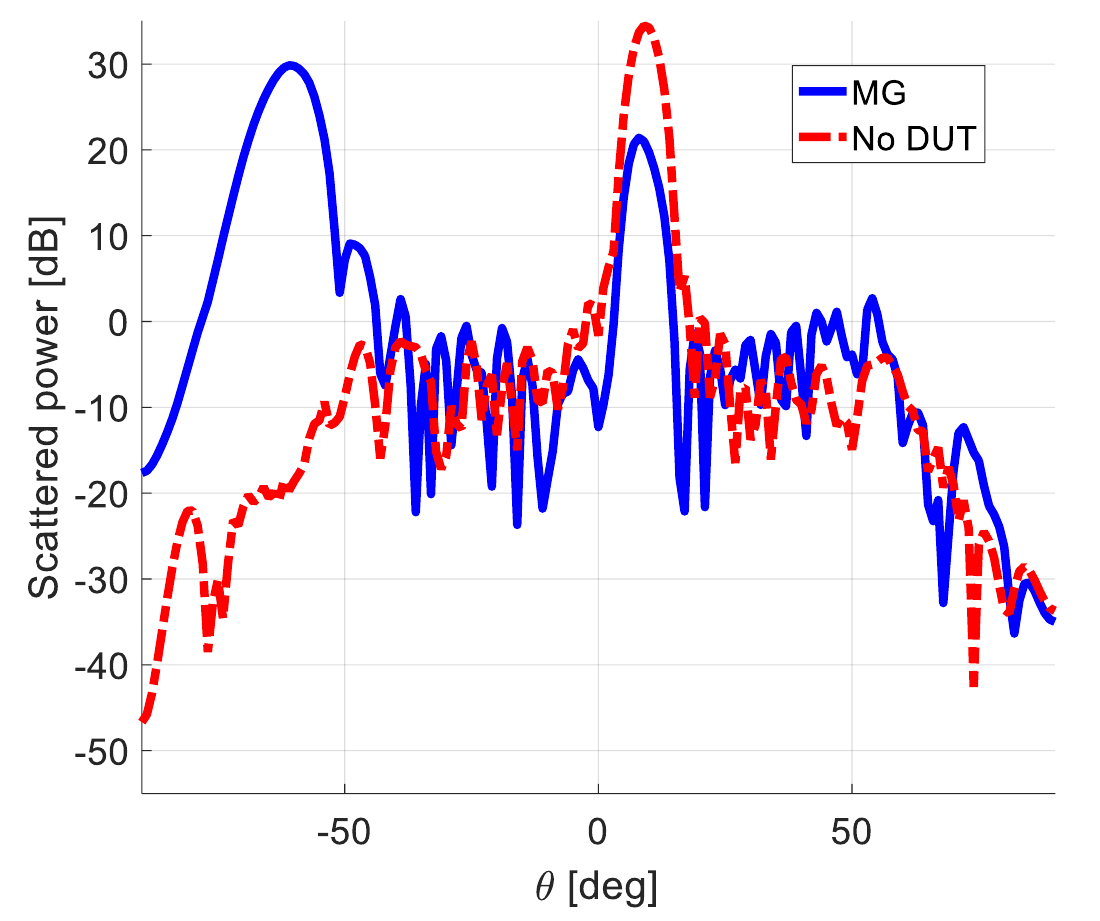}
\caption{Experimentally recorded forward ($z>0$) scattering pattern at the operating frequency of $f=20\mathrm{GHz}$. The received power as a function of $\theta$ as obtained when the MG board was excited (solid blue) is compared to the reference pattern recorded in the absence of the DUT (dash-dotted red).}
\label{Fig:Exp_scattering_pattern}
\end{figure}

Figure \ref{Fig:Exp_scattering_pattern} compares the measured forward scattering pattern produced by the illuminated DUT with the reference one (direct illumination in the absence of the MG) at the designated operating frequency of $f=20\mathrm{GHz}$.
As predicted, the MG prototype suppresses the direct mode in transmission ($m=0$) observed at $\theta_\mathrm{in}=10^\circ$, redirecting the incident power towards the anomalous refraction mode ($m=-1$) at $\mathrm{\theta_{out}}=-60^{\circ}$. 


\begin{figure}[t]
\centering
\includegraphics[width=2.5in]{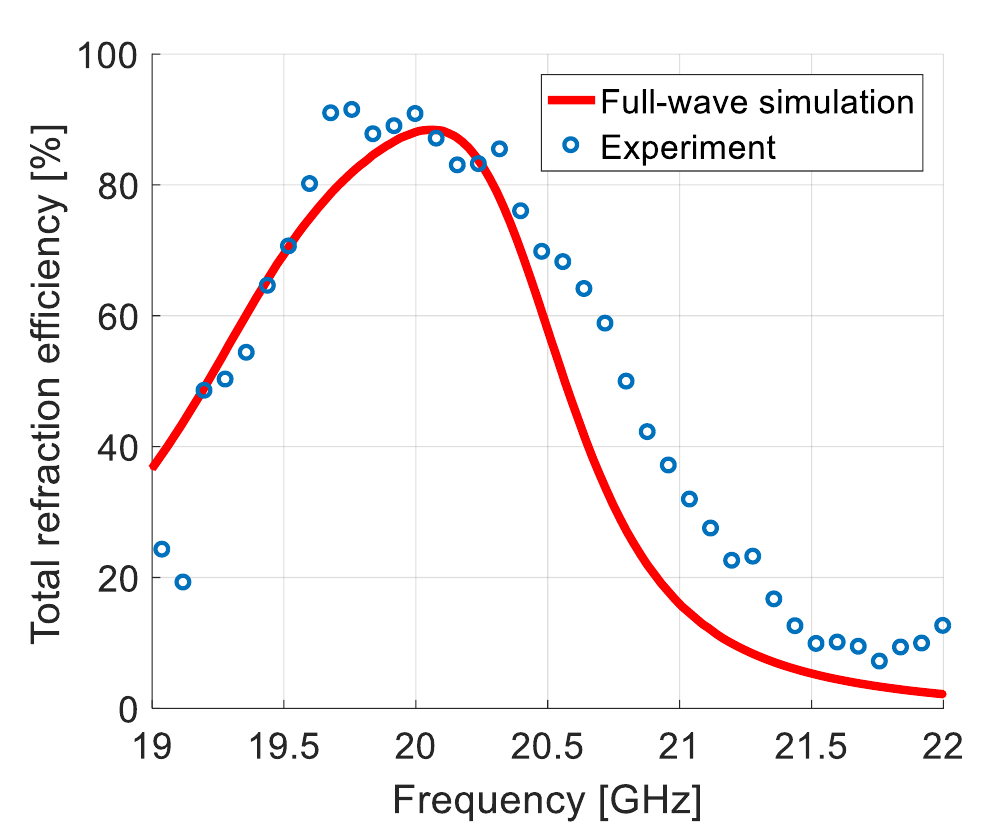}
\caption{The total anomalous refraction efficiency of the prototype MG as function of frequency. The experimental results (blue circles) are compared with the results obtained via full-wave simulation (solid red).}
\label{Fig:Exp_efficiency}
\end{figure}

To better assess the agreement between the theoretical design and the fabricated prototype, we compare in Fig. \ref{Fig:Exp_efficiency} the \emph{overall} anomalous refraction efficiency $\eta_\mathrm{tot}$ (taking into account both spurious scattering and absorption) as a function of frequency, as obtained from full-wave simulations (red solid line) and from the experimental data (blue circles). The latter is evaluated from the peak measured gain values for the MG scattering pattern $G_{\mathrm{MG}}\left(\theta_\mathrm{out}\right)$ and the bare Gaussian beam excitation $G_\mathrm{direct}\left(\theta_\mathrm{in}\right)$, applying the suitable effective aperture size normalization as in \cite{wong2018perfect,wong2018binary,diaz2017generalized,asadchy2017flat}, namely, 
\begin{equation}
\label{eq:Gain_calc_exp}
\begin{aligned}
\eta_{\mathrm{tot}}=\frac{G_{\mathrm{MG}}\left(\theta_\mathrm{out}\right)}{G_{\mathrm{direct}}\left(\theta_\mathrm{in}\right)}\cdot \frac{\cos\theta_\mathrm{in}}{\cos\theta_\mathrm{{out}}}
\end{aligned}
\end{equation}
As opposed to our previous experimental work \cite{rabinovich2019experimental}, the near-field measurement system used herein allowed accurate low-noise evaluation of the far-field patterns (time gating included), facilitating the use in this methodology. 

The results as presented in Fig. \ref{Fig:Exp_efficiency} reveal an excellent agreement between simulated and measured refraction efficiencies, verifying once again the fidelity of our analytical model and design scheme. In terms of achieved performance, the plots indicate that as much as $90\%$ of the incident power at $\theta_\mathrm{in}=10^\circ$ is coupled to the anomalous refraction mode at $\theta_\mathrm{out}=60^\circ$, confirming the ability of multilayered multielement MGs to manipulate beams in transmission with very high efficiency. In fact, previous realizations of reflectionless wide-angle beam refraction at microwave frequencies with omega-bianisotropic metasurface prototypes \cite{lavigne2018susceptibility,chen2018theory} yielded lower efficiencies, despite the significantly simpler physical structure offered by the MG devised herein.


\subsection{Non-local lens array}
\label{subsec:lens}
\begin{figure}[t]
\centering
\includegraphics[width=2.5in]{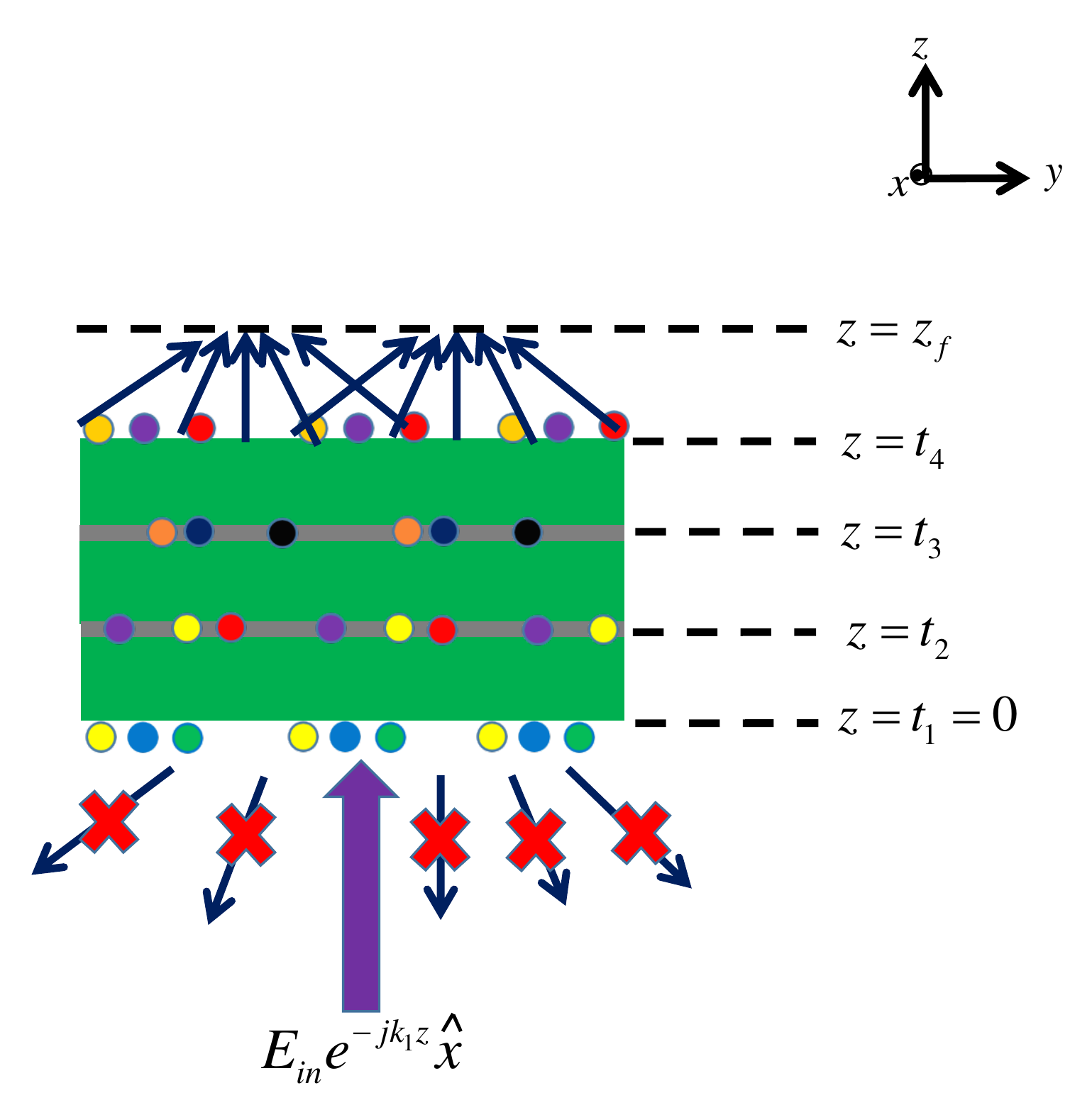}
\caption{Physical configuration of the non-local lens array MG prototype, featuring 12 meta-atoms arranged in 4 layers. Schematically, green layers denote the Rogers RO3003 laminates, whereas the thin grey slabs mark the Rogers 2929 bondply between laminates.
}
\label{Fig:Config_Lens}
\end{figure}

As stated in Section \ref{Sec:Introduction}, in addition to demonstrating the efficacy and fidelity of the analytical model in designing highly-efficient MGs for controlling both reflected and transmitted FB modes, we wish to show that the same methodology can manipulate not only the power partition but also the phase of individual FB modes. This is especially useful for focusing applications, where different rays should be delayed by different phases, such that a focus will be formed. Considering the periodic nature of the MGs discussed herein, we utilize the formalism developed in Section \ref{sec:Theory} to design a $\Lambda=2.1\lambda$ periodic lens array, focusing the normally incident power ($\theta_\mathrm{in}=0$) at $f=20\mathrm{GHz}$ into a series of foci $(y_\nu,z)=(y_{\mathrm{f}}+\nu\Lambda,z_{\mathrm{f}})$, where $y_{\mathrm{f}}=0.61\lambda$ and $z_{\mathrm{f}}=3.31\lambda$ away from the MG plane (Fig. \ref{Fig:Config_Lens}). 

Interestingly, as opposed to conventional flat lenses \cite{hasman2003polarization,lin2014dielectric,aieta2015multiwavelength,chen2018broadband}, including the ones recently demonstrated with MGs \cite{paniagua2018metalens, neder2019combined}, the focusing MG presented herein is not designed by locally deflecting the incident rays towards the focus via modification of the anomalous refraction angle (varying the local phase gradient). Instead, the devised lens array is of non-local nature, from two related perspectives: first,
the power contributing to each transmitted FB mode is coupled from the incident rays via near-field interactions, implicitly involving power transfer via surface waves along the MG plane as resulting from the synthesis scheme; second, 
rays contributing to a certain focus often depart from neighbouring periods (Fig. \ref{Fig:Config_Lens}), enabling multiple sharp (diffraction-limited) focal spots with reduced aperture size due to this shared aperture effect, increasing the effective aperture efficiency per focus.

As per \eqref{eq:num_prop_modes}, the lens array periodicity gives rise $2M=10$ propagating FB modes ($5$ in reflection and $5$ in transmission). Consequently, we choose the number of meta-atoms per period to be $K=12\geq10$, evenly distributed in the interfaces between three stacked Rogers RO3003 laminates ($N=5$, $\varepsilon_2=\varepsilon_3=\varepsilon_4=3\varepsilon_0$, and $\tan\delta=0.001$), with 3 elements in each of the 4 interfaces. As in Section \ref{subsec:refractor}, each of the laminates is of thickness $60 \mathrm{mil}=1.52 \mathrm{mm}$, and we bond them using $2 \mathrm{mil}=0.051 \mathrm{mm}$ thick Rogers 2929 bondply ($\varepsilon=2.94$ and $\tan \delta= 0.003$) to form a $0.311\lambda=4.67 \mathrm{mm}$ multilayer PCB (Fig. \ref{Fig:Config_Lens}). Once again, we are required to reduce the number of nonlinear constraints in \eqref{eq:passivity_condition} and \eqref{eq:lossless_condition} to compensate for the limited set of laminate thicknesses commercially available, as detailed in Appendix \ref{app:optimization}. Nonetheless, as shall be shortly demonstrated, even with the compacted set of constraints, achieved by grouping adjacent meta-atoms and applying the constraints on them collectively, the Matlab program solving this nonlinear problem (Section \ref{subsec:passivity}) yields MG lens arrays with very good performance.

To initiate the design procedure, we define the desired modal \emph{complex} amplitudes such that the foci will be formed in the prescribed position. Following simple geometrical optics consideration \cite{born1999principles} we demand that the $5$ transmitted modes would carry equal power, while their phases differ according to the angle of propagation as to arrive at the same phase to the focal point. In addition, we would ideally want no reflections from the lens array; thus, we require all 5 propagating reflected modes to vanish. Explicitly, we set $E_{m}^\mathrm{ref}=0$ and $E_{m}^\mathrm{trans}=E_\mathrm{in}\sqrt{\beta_{0,1}/\left(5\beta_{m,1}\right)}\exp\left\{j\beta_{m,1}z_f+j k_{t,m}y_{f}\right\}$ for $m=-2,-1,0,1,2$, in consistency with \eqref{eq:Tot_power_conservation}.

\begin{table*}[t]
\centering
\begin{threeparttable}[b]
\renewcommand{\arraystretch}{1.3}
\caption{Design specifications of the non-local lens array MG (corresponding to Fig. \ref{Fig:Config_Lens}).
}
\label{tab:Lens}
\centering
\begin{tabular}{|c||c|c|c||c|c|c||c|c|c||c|c|c|}
\hline \hline

& \multicolumn{3}{c||}{1st layer}  &  \multicolumn{3}{c||}{2nd layer} &  \multicolumn{3}{c||}{3rd layer} &  \multicolumn{3}{c|}{4th layer} \\	\hline
& $k=1$ & $k=2$ & $k=3$ 
	& $k=4$ & $k=5$ & $k=6$ 
		& $k=7$ & $k=8$ & $k=9$ 
			& $k=10$ & $k=11$ & $k=12$ \\ \hline \hline
$h_k [\lambda]$ 
& $0$ & $0$ & $0$
	& $0.101$ & $0.101$ & $0.101$
		& $0.206$ & $0.206$ & $0.206$
			& $0.311$ & $0.311$ & $0.311$   \\	\hline
$d_k [\lambda]$ 
& $0$ & $-0.716$ & $-0.298$
	& $-0.959$ & $0.710$ & $0.428$
		& $0.097$ & $-0.276$ & $-0.718$
			& $0.361$ & $0.944$ & $-0.782$   \\	\hline
$W_k [\mathrm{mm}]$ 
& $1.8718$ & $0.791$ & $1.1156$
	& $1.0978$ & $1.0632$ & $1.2468$
		& $1.7016$ & $1.875$ & $1.4885$
			& $2.3751$ & $3.2802$ & $2.6192$  \\	\hline 
\hline 
\end{tabular}
\end{threeparttable}
\end{table*}

%
%
%
%
%

Similar to Section \ref{subsec:refractor}, we substitute these modal field amplitudes into the RHS of \eqref{eq:Matrix_equation}, which allows, for given $(d_k, h_k)$, resolution of the system of $2M=10$ equations and $K=12$ unknowns using a Moore-Penrose pseudoinverse method, yielding the currents $I_k$ that are to be induced on the wires to obtain the prescribed diffraction. This forward problem formulation is used in MATLAB's \texttt{lsqnonlin} to minimize deviations from the passivity conditions \eqref{eq:passivity_condition} and to reduce absorption \eqref{eq:lossless_condition} using the available degrees of freedom, namely, the meta-atom respective offsets (see Appendix \ref{app:optimization}) 
Once the optimal positions $(d_k,h_k)$ are obtained with the corresponding $I_k$, required load impedances and respective capacitor widths can be assessed as detailed in Section \ref{subsec:Coupling}. The optimal parameters resulting from this process for the lens array specifications prescribed herein are listed in Table \ref{tab:Lens}. 

\begin{figure}[t]
\centering
\includegraphics[width=2.2in]{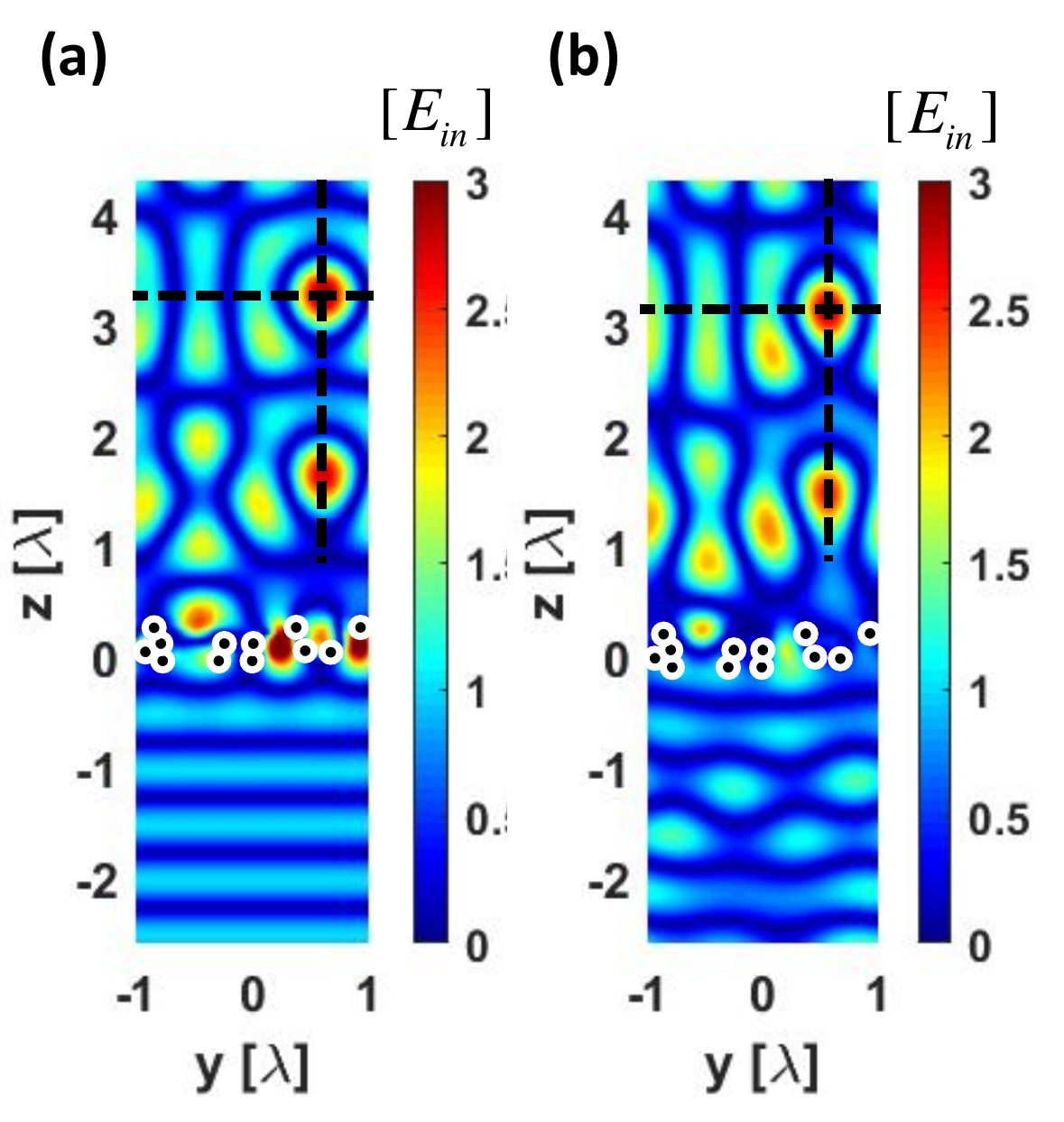}
\caption{Electric field distribution $|\Re\{E_{x}(y,z)\}|$ corresponding to the non-local lens array MG prototype when excited from $\mathrm{\theta_{in}}=0^{\circ}$, (a) as predicted by the analytical model at the design frequency $f=20 \mathrm{GHz}$ and (b) as recorded by full-wave simulations at the optimal operating frequency $f=19.9 \mathrm{GHz}$. A single period of the MG is presented; white circles denote the positions of the meta-atoms, and dashed lines mark the coordinates of the designated focal point $\left(y_\mathrm{f},z_\mathrm{f}\right)$.}
\label{Fig:Field_Lens}
\end{figure}

In order to validate our theoretical synthesis scheme for the case of the lens array, we defined the devised MG configuration in a full-wave solver (CST Microwave Studio) and simulated it under periodic boundary conditions. A comparison between the electric field distribution as predicted by the analytical model and as recorded in full-wave simulation is presented in Fig. \ref{Fig:Field_Lens} (white circles denote the meta-atom constellation). The field distribution derived using the analytical model [Fig. \ref{Fig:Field_Lens}(a)] is evaluated in the designed operating frequency of $f=20\mathrm{GHz}$, while the one extracted from the full-wave simulation operates at $f=19.9\mathrm{GHz}$, in which minimum reflection was observed (we consider this to be the optimal actual operating frequency from now on). Such minor frequency shifts ($0.5\%$) are quite reasonable when comparing analytical results with realistic simulations of a fabrication-ready layout, especially if we consider the fact that no full-wave optimization was involved in the synthesis procedure.
Indeed, except for the expected mild discrepancies in the close vicinity of the meta-atoms \cite{epstein2017unveiling,rabinovich2018analytical,rabinovich2019experimental}, very good correspondence is observed (note the clear focus at the designated coordinates), confirming the theoretical design. 

\begin{figure}[t]
\centering
\includegraphics[width=2.5in]{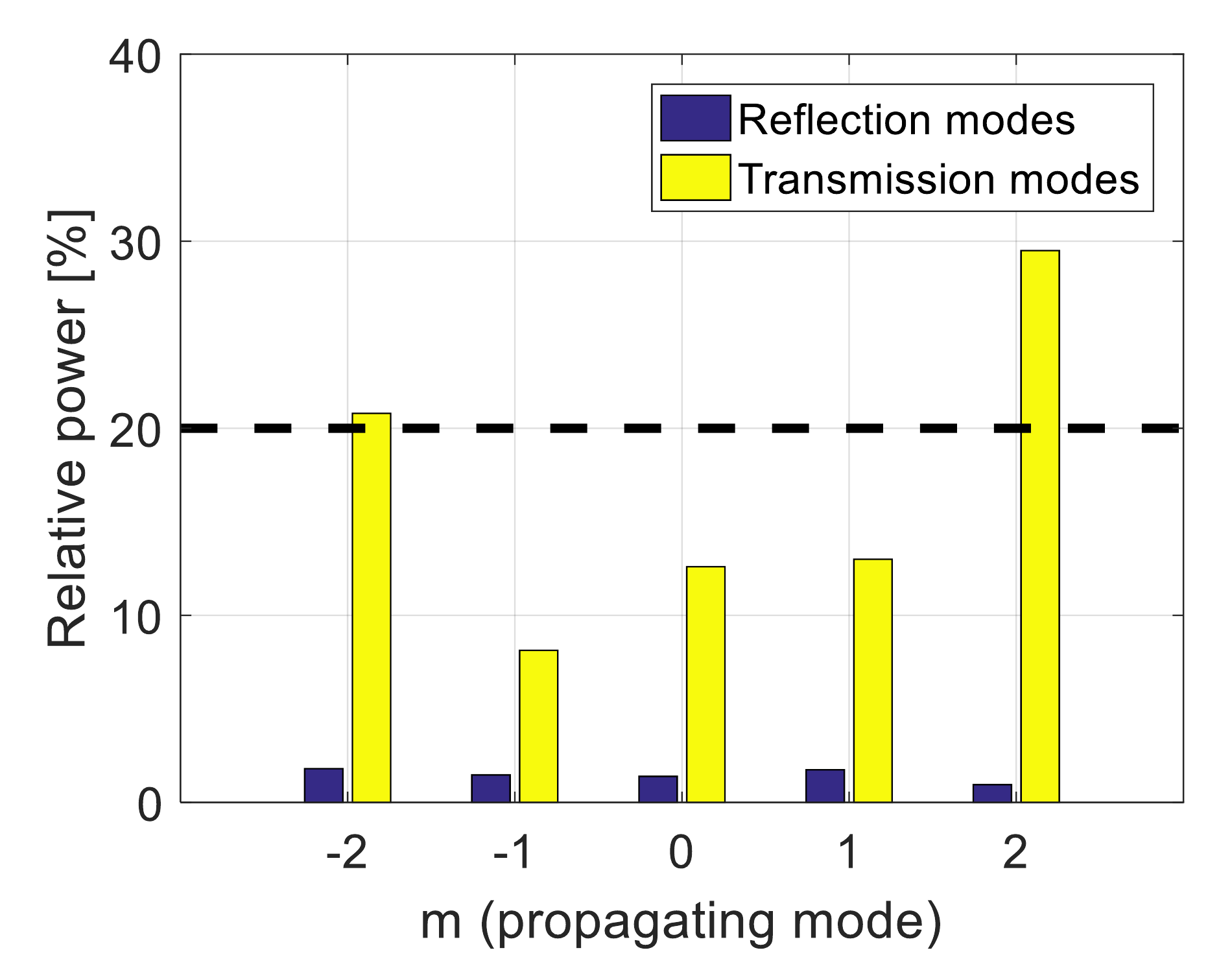}
\caption{Relative power coupled to the various reflected (blue bars) and transmitted (yellow bars) propagating FB modes, as obtained from full-wave simulations at the optimal operating frequency $f=19.9\mathrm{GHz}$. The dashed horizontal line marks the ideal power partition, with $20\%$ of the incident power coupled to each of the transmitted mode.}
\label{Fig:Bar_lens}
\end{figure}


Quantitatively, we can assess the designed MG performance in two manners. First, in the far field, the lens array acts as a forward diffuser, designed for minimal reflections and equal partition of power between the manifold of forward propagating modes. Correspondingly, we present in 
Fig. \ref{Fig:Bar_lens} the relative power coupled to the various modes in reflection and transmission, as obtained from full-wave simulations at the optimal operating frequency. As can be observed in Fig. \ref{Fig:Bar_lens}, although reflections do not vanish completely, they are drastically suppressed, with less than $2\%$ coupled to each of the reflected modes, and only $7.4\%$ reflections overall. Larger deviations are observable in the transmitted modal amplitudes with respect to the design goal of equipower partition, according to which $20\%$ of the incident power should be coupled to each of the transmitted modes (dashed line).

The main reason for this deviation is the fact that while the analytical model produces the design assuming passive and lossless meta-atoms, this goal, manifested by \eqref{eq:passivity_condition} and \eqref{eq:lossless_condition}, cannot be perfectly achieved. As previously discussed in Section \ref{subsec:refractor}, limiting ourselves to standard commercially available laminate thicknesses reduces the number of available degrees of freedom (see Appendix), which, in turn, diminishes the effectiveness of the MATLAB optimization. Consequently, the design specifications may include small active/lossy components in the optimal load impedances, which, when realized using passive realistic printed capacitors, cause some deviations from the synthesis goals. Similarly, while we attempt to minimize absorption as much as possible, the induced currents in the final configuration do incur certain losses. Although we may estimate their extent quantitatively [by summing the contributions \eqref{eq:lossless_condition}], we cannot predict how they would affect the overall performance. In the designed MG, for instance, the model predicted that $5\%$ of the incident power would be absorbed in the MG, which, although comparable with the value of $8.35\%$ absorption obtained from full-wave simulations, may deter the power partition.
Nonetheless, as can be clearly seen from Fig. \ref{Fig:Field_Lens}, the overall performance of the intricate 4-layer 12-elements MG lens array is not deteriorated by the slight amplitude 
offset, forming the foci very close to the desired positions with an overall high transmission efficiency of $84\%$ (accounting for both absorption and reflection).

\begin{figure}[t]
\centering
\includegraphics[width=2.5in]{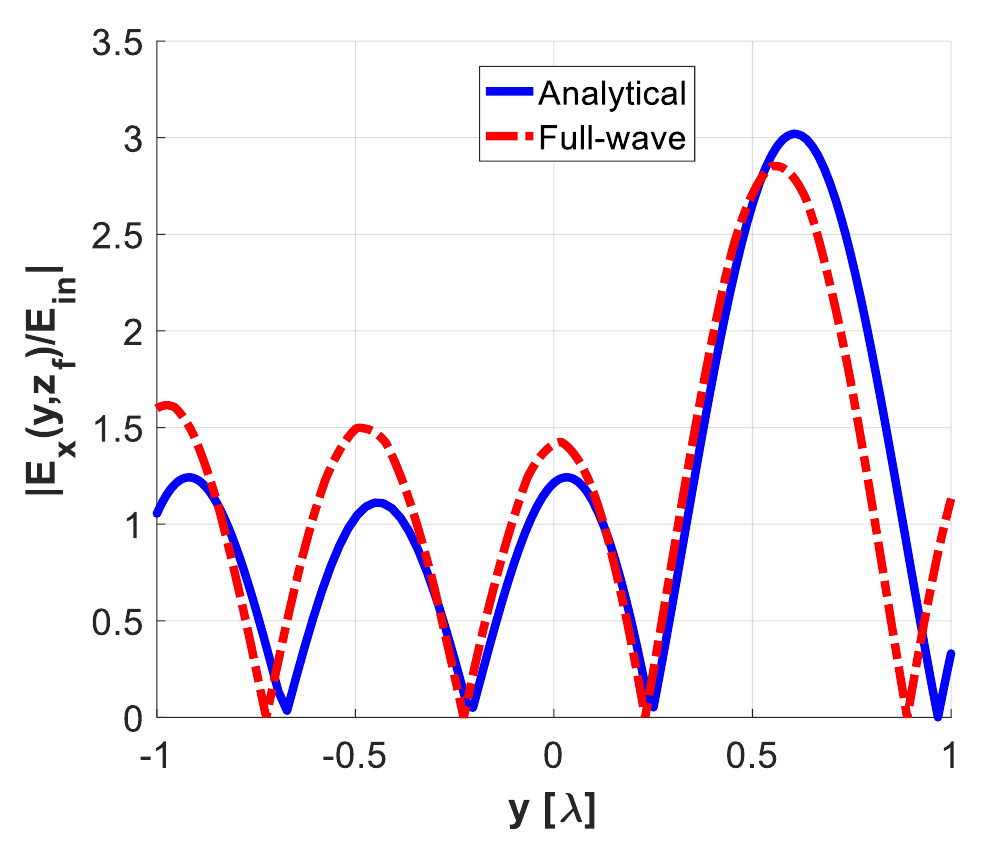}
\caption{Electric field lateral magnitude profile $|E_{x}(y,z_\mathrm{f})|$ at the focal plane $z=z_\mathrm{f}$ as predicted by the analytical model at the design frequency $f=20\mathrm{GHz}$ (blue solid line) and by full-wave simulations at the optimal operating frequency $f=19.9\mathrm{GHz}$ (red dashed line).}
\label{Fig:Focus_y}
\end{figure}

\begin{figure}[t]
\centering
\includegraphics[width=2.5in]{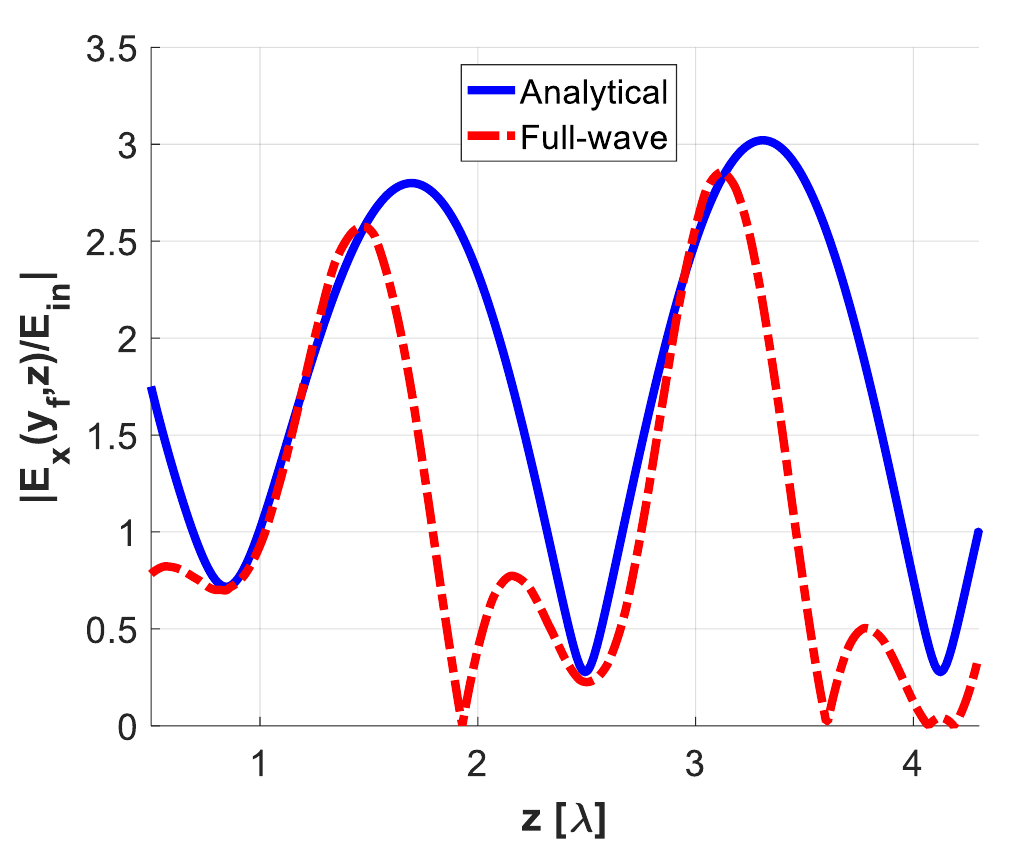}
\caption{Electric field longitudinal magnitude profile $|E_{x}(y_\mathrm{f},z)|$ through the focal point, as predicted by the analytical model at the design frequency $f=20\mathrm{GHz}$ (blue solid line) and by full-wave simulations at the optimal operating frequency $f=19.9\mathrm{GHz}$ (red dashed line).}
\label{Fig:Focus_z}
\end{figure}

\begin{figure}[t]
\centering
\includegraphics[width=3.4in]{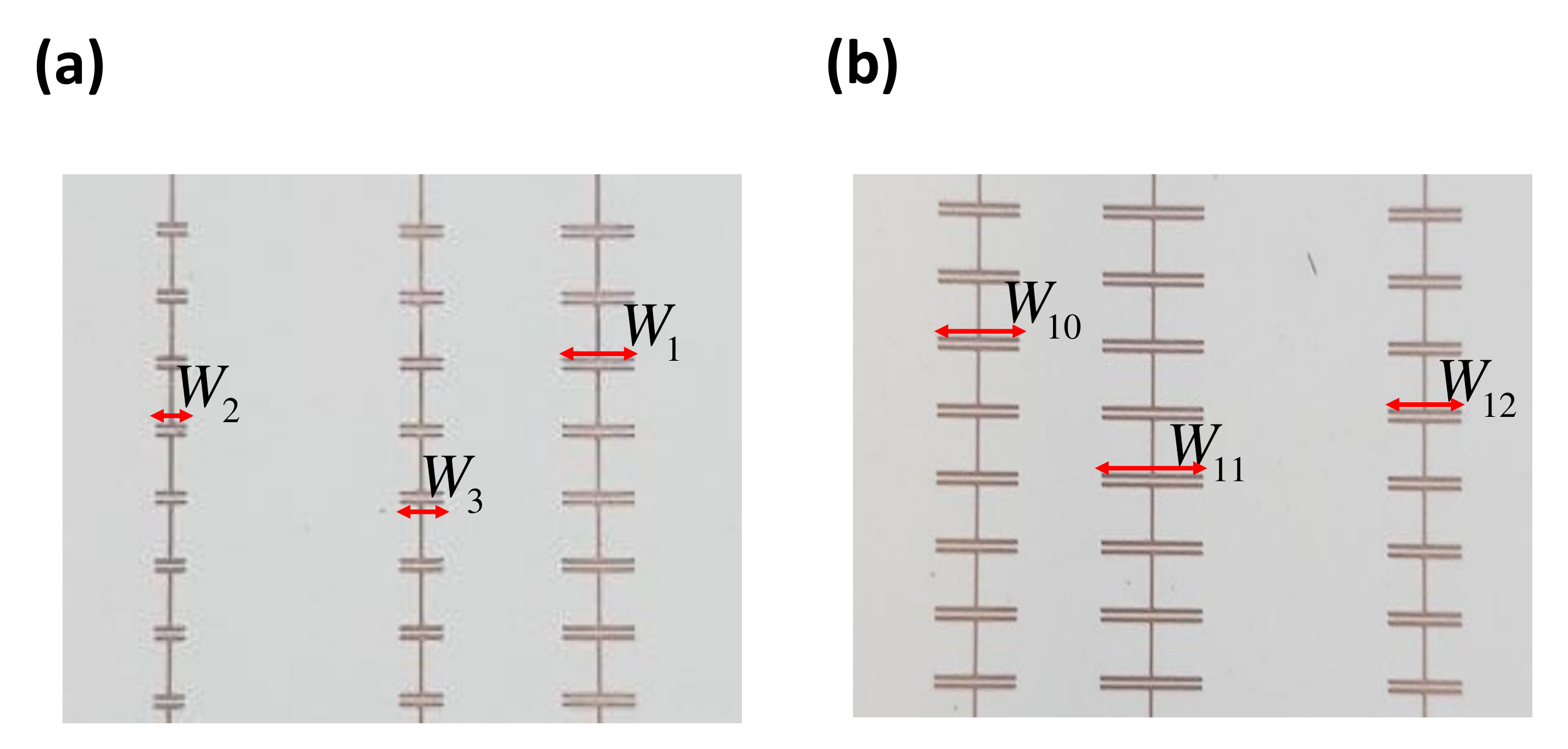}
\caption{Zoom-in on one lateral and eight longitudinal periods of the fabricated MG non-local lens prototype (a) top view; (b) bottom view. Loaded wires numbered ${k=4,...,9}$ are embedded in the internal layers of the PCB, and thus are not shown; capacitor widths correspond to Table \ref{tab:Lens}.}
\label{Fig:Zoom_in_board_Lens}
\end{figure}



\begin{figure*}[t]
\centering
\includegraphics[width=7.2in]{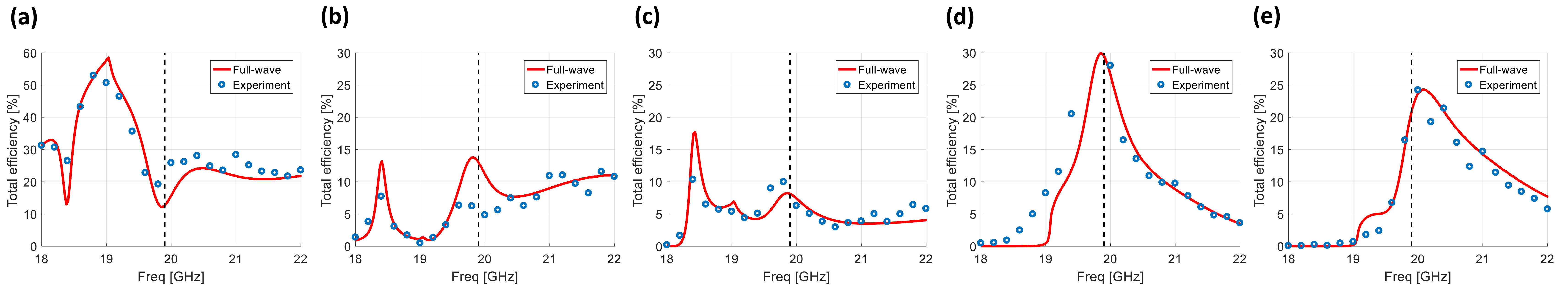}
\caption{The portion of incident power coupled to the various propagating FB modes in transmission as a function of frequency, for the non-local lens array MG prototype. The experimental results (blue circles) are compared with the results obtained via full-wave simulation (solid red) for the relevant mode indices (a) $m=0$, (b) $m=1$, (c) $m=-1$, (d) $m=2$, and (e) $m=-2$. Dashed vertical lines denote the actual optimal operating frequency $f=19.9\mathrm{GHz}$.}
\label{Fig:Modes_FF_EXP}
\end{figure*}

These observations are further supported by a detailed examination of the electric field magnitude profiles along the $y$ axis (Fig. \ref{Fig:Focus_y}) and $z$ axis (Fig. \ref{Fig:Focus_z}) intersecting at the designated focus $(y_{\mathrm{f}},z_\mathrm{f})$, denoted by dashed horizontal and vertical lines, respectively, in Fig. \ref{Fig:Field_Lens}. Figure \ref{Fig:Focus_y} reveals that the analytically predicted fields (solid blue) at the focal plane $\left|E_x\left(y,z_f\right)\right|$ agree very well with the ones retrieved from full-wave simulations of the physical structure at the actual optimal operating frequency (dashed red). The deviation from the expected horizontal position of the peak $y_{f}$ is less than $7\%$, with both results indicating a diffraction limited full-width half maximum (FWHM) of roughly $0.46\lambda$. We note that such a tight focus could only be achieved due to the fact that all periods of the non-local MG lens array participate in the formation of foci, with some of the power contributing to the focus arriving from neighbouring periods (Fig. \ref{Fig:Config_Lens}). 

\begin{figure}[t]
\centering
\includegraphics[width=3in]{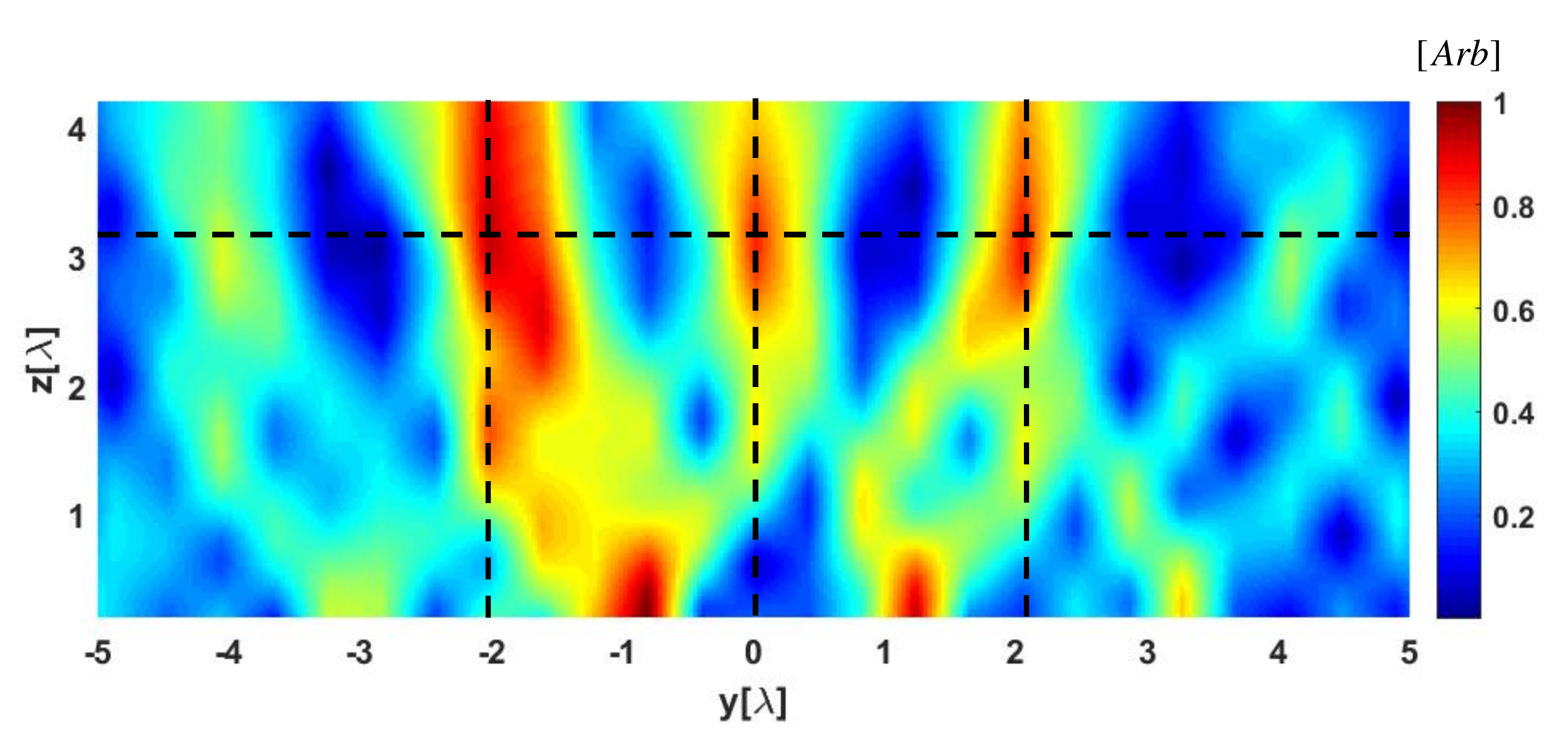}
\caption{Electric field distribution $|\Re\{E_{x}(y,z)\}|$ corresponding to the non-local lens array MG prototype, as obtained from measurements at the optimal operating frequency $f=19.9 \mathrm{GHz}$. Five lateral periods are presented, in view of the finite illumination area of the Gaussian beam excitation. Horizontal black dashed line mark the expected focal plane $z=z_\mathrm{f}^\mathrm{sim}$, and vertical black dashed lines denote the lateral positions of the formed foci. }
\label{Fig:Lens_2D_EXP_SIM_comp}
\end{figure}

\begin{figure}[t]
\centering
\includegraphics[width=3in]{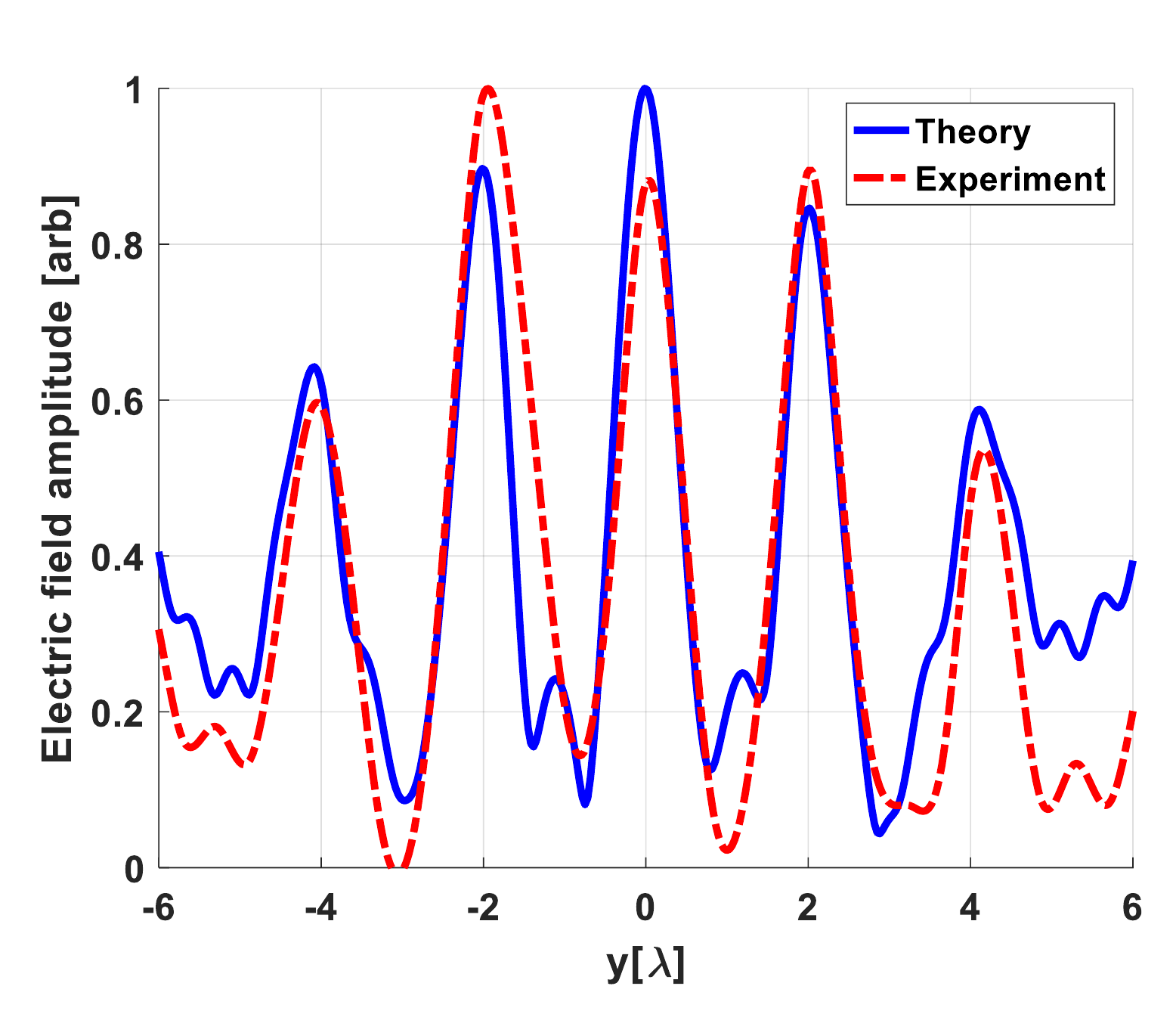}
\caption{Electric field magnitude profile $|E_{x}(y,z_\mathrm{f})|$ at the focal plane $z=z_\mathrm{f}^\mathrm{sim}$ in the optimal operating frequency $f=19.9\mathrm{GHz}$, as predicted by the theoretical calculation relying on the analytical model and considering the finite illumination area of the Gaussian beam (blue solid line, beam diameter $5\lambda$@$z=0.43\lambda$) and as extracted from the experimental data (red dashed line).}
\label{Fig:Lens_1D_EXP_Vs_y}
\end{figure}
Analogously, the variation of the electric field magnitude as one moves vertically away from the MG along the lateral focus coordinate $\left|E_x\left(y_{f},z\right)\right|$, presented in Fig. \ref{Fig:Focus_z}, indicates good correspondence between the analytical model (solid blue) and full-wave simulations (dashed red). The intended focus (around $z=z_f=3.31\lambda$) is accompanied in both profiles by a secondary focus (around $z=1.8\lambda$), appearing due to unintentional constructive interference of the phase-shifted FB modes. As in Fig. \ref{Fig:Focus_y}, the field variation trend is similar in both profiles, with about $7\%$ deviation in the position of the focus, evaluated from simulations to be formed at $z=z_\mathrm{f}^\mathrm{sim}=3.11\lambda$. 

With these encouraging full-wave verification results, we proceeded to experimentally validate our design. The design specifications listed in Table \ref{tab:Lens}, obtained based on our analytical model, were used to define a multilayer PCB layout (Fig. \ref{Fig:Config_Lens}) containing $8$ periods along the $y$ axis and $169$ periods along the $x$ axis, fabricated on a $9''\times12''$ board (Fig. \ref{Fig:Zoom_in_board_Lens}). The experimental setup used for characterizing the non-local lens array is exactly the same as the one described in Section \ref{subsec:refractor} and depicted in Fig. \ref{Fig:Exp_setup}, with the only difference being the alignment of the Gaussian beam antenna with respect to the DUT, rotated to form an incident angle of $\theta_\mathrm{in}=0$ as required. Due to mechanical setup constraints, the scan plane remained at $z=16.5\mathrm{cm}\approx11\lambda$, covering the same area of $1.4\mathrm{m}\times0.28\mathrm{m}$, with far and near field evaluation following standard equivalence principle and back-projection techniques, respectively, implemented (including probe correction) by the MVG/Orbit-FR MiDAS Measurement Software Suite.

As with the full-wave validation, we use both far and near field data to characterize and verify the prototype performance experimentally. The far-field measurement yields the coupling efficiency to each of the $5$ FB propagating modes in transmission ($m=-2,-1,0,1,2$), evaluated following \eqref{eq:Gain_calc_exp}, with the suitable modal peak gain values and corresponding angles $\theta_{\mathrm{out},m}=\arcsin(k_{t,m}/k_1)$. Figure \ref{Fig:Modes_FF_EXP} compares the coupling efficiency to the various modes as recorded in the experiment (blue circles) with the one obtained from full-wave simulations (red solid line), as a function of frequency. Very good correspondence between the two is clearly observed across all $5$ forward propagating modes, validating the fabricated MG functionality in terms of power partition. 

To verify also the near-field characteristics of the MG non-local lens array, we back-projected the measured complex field amplitudes to create a map of the electric field magnitude distribution $|E_{x}(y,z)|$ at the actual optimal operating frequency $f\approx19.9\mathrm{GHz}$ (Fig. \ref{Fig:Lens_2D_EXP_SIM_comp}). Three dominant foci can be observed at the expected focal plane $z=z_\mathrm{f}^\mathrm{sim}=3.11\lambda$ (horizontal black dashed line), with the lateral coordinates of the focal points evaluated, from left to right, as $y_{\mathrm{f},-1}^\mathrm{exp}=-1.98\lambda$, $y_{\mathrm{f},0}^\mathrm{exp}=0.03\lambda$, and $y_{\mathrm{f},1}^\mathrm{exp}=2.03\lambda$, separated on average by $2.02\lambda$. The correspondence with the designated focal plane, as well as the close proximity of the lateral foci separation to the MG periodicity $\Lambda=2.1\lambda$, indicate an overall good agreement with the theoretical predictions. The secondary foci at $z\approx1.8\lambda$ denoted in the simulated results (Fig. \ref{Fig:Field_Lens}) are observed as well at the expected lateral coordinates, but with lower magnitude.

To further assess the properties of the experimentally observed foci in view of the predicted results, 
we wish to theoretically estimate the field profile at the focal plane while taking into account the finite size of the produced MG board and Gaussian beam illumination. 
Fortunately, this can be readily performed by using the fields at the MG aperture, as evaluated during the design process from our analytical formulation (Section \ref{sec:Theory}). Specifically, we utilize the fields calculated for the infinite periodic MG lens array \eqref{eq:Total_electric_field_space} and duplicate them along $8$ lateral periods, matching the size of the fabricated board. Thence, we use a Gaussian envelope as a window function to these aperture fields, to approximate the fields more realistically formed on the MG aperture by the finite-diameter Gaussian beam excitation. Finally, the equivalence principle is used to advance these fields to the focal plane $z=z_\mathrm{f}^\mathrm{sim}$ (horizontal black dashed line in Fig. \ref{Fig:Lens_2D_EXP_SIM_comp}), for comparison with the measured results $\left|E_x\left(y,z_\mathrm{f}^\mathrm{sim}\right)\right|$ \cite{balanis2012advanced}.

The result of this calculation is presented in Fig. \ref{Fig:Lens_1D_EXP_Vs_y}, with the theoretical fields at $z=0.43\lambda$ taken as aperture fields and a Gaussian beam with a diameter of $5\lambda$ (where $85\%$ of the power is concentrated) used as a window function\footnote{The measured Gaussian beam diameter at $z=0$ was $7.7\lambda$. We associate the difference between this value and the effective Gaussian window diameter at $z=0.43\lambda$ best fitting the experimental MG near-field results with the non-local nature of the MG lens array, which induces power flow along the $y$ axis that may cause beam narrowing.}. As can be clearly seen, these parameters yield very good correspondence between the field profile observed in the experiment, and the one anticipated for the finite MG prototype and Gaussian beam illumination. Importantly, this comparison clarifies the main reason for the reduced focus sharpness, with FWHM recorded in the experiment reaching $\approx 0.87\lambda$. Indeed, this focus broadening can be traced back to the reduced effective number of periods illuminated by the Gaussian beam, preventing from all the modes to contribute to the focus formation (see Fig. \ref{Fig:Config_Lens} and discussion at the beginning of this Subsection); the predicted FWHM for the central focus can be restored, however, if more periods are effectively illuminated by the Gaussian beam. 

Overall, considering the combined far and near field evidence, and possible inaccuracies due to minor fabrication discrepancies, alignment errors, and back-projection and probe correction post-processing techniques, the presented results unambiguously verify the performance of the proposed non-local lens array, highlighting the potential of MGs to control not only magnitude but also phase of scattered modes, yielding devices with complex functionalities.


\section{Conclusion}
To conclude, we have presented a general and rigorous analytical model for the design of multilayered multielement capacitively-loaded wire MGs for arbitrary control over the amplitude and phase of a large number of diffraction modes, reflected and transmitted alike. Formulating the design constraints analytically following this model allows proper identification of the required number of degrees of freedom and judicious choice of the number of meta-atoms per period necessary to meet the design goals. Based on this formulation, we have devised an efficient synthesis scheme for such MGs, prescribing the detailed multilayer PCB layout up to the complete copper trace geometry, without as much as a single optimization in a full-wave solver.

The fidelity of the proposed scheme, incorporating measures to yield practical passive designs with reduced conductor loss, was demonstrated and verified using full-wave simulations and laboratory measurements for two different prototypes operating at K band ($f=20\mathrm{GHz}$). First, a 2-layer 4-element MG implementing perfect anomalous refraction has been designed, simulated, fabricated, and characterized, showing excellent agreement between theory and experiments. Specifically, $\approx90\%$ of the Gaussian beam power impinging at $\theta_\mathrm{in}=10^\circ$ has been successfully coupled to the $m=-1$ FB mode in transmission ($\theta_\mathrm{out}=-60^\circ$), showing how MGs can be systematically designed to manipulate beams, not only in reflection, as has been demonstrated before, but also in transmission. 
In the second example, a 4-layer 12-element non-local lens array was designed, fabricated and verified, practically eliminating the 5 spurious reflected propagating FB modes while exercising control over both phase and amplitude of the 5 transmitted FB modes. This efficient manipulation of the modal phase of numerous transmitted modes, demonstrated for MGs for the first time, to the best of our knowledge, allows formation of sharp periodic foci across a prescribed focal plane, as verified by both simulations and measurements.

Overall, the general theoretical treatment and comprehensive experimental validation form a flexible and reliable framework for systematic full-wave-optimization-free synthesis of MGs for arbitrary diffraction control. By harnessing the general multilayer multielement PCB configuration, the proposed methodology offers immense versatility, enabling power distribution and wavefront shaping in both reflection and transmission. As implied by the non-local lens array demonstration, we expect that such powerful and efficient engineering tools, based on a rigorous analytical model, would facilitate the development of new intricate MG-based devices with unconventional properties.

\section*{acknowledgment}
This research was supported by the Israel Science Foundation (Grant No. 1540/18). 
The authors also wish to thank the team of MVG/Orbit-FR in Israel for continuous technical support regarding acquisition and analysis with the planar near-field measurement system. In addition, they thank Yuri Komarovsky, Ben-Zion Joselson, and Denis Dikarov of the Communication Lab at the Technion for their assistance and advice with regards to the experimental setup.

\appendices
\section{Enforcing the constraints for passivity and loss minimization}
\label{app:optimization}
In Section \ref{subsec:refractor}, we stressed the difficulty associated with the limited variety of commercially available laminates with which we assemble our MG prototypes. In the context of the synthesis methodology, this limitation effectively reduces the number of degrees of freedom we may use to meet the design goals, as the vertical offsets of the various meta-atoms $h_k$ can no longer be chosen arbitrarily. Therefore, the number of constraints dictated by \eqref{eq:passivity_condition} and \eqref{eq:lossless_condition} is larger than the available free parameters, which introduces difficulties in utilizing nonlinear-constraint solver \texttt{lsqnonlin} by MATLAB, chosen for resolving the MG specifications (Section \ref{sec:Theory}). 
Practically, we can overcome this problem by forming a compacted set of constraints for passivity and loss minimization out of the full set formulated in \eqref{eq:passivity_condition} and \eqref{eq:lossless_condition}, matching the reduced number of degrees of freedom, without compromise on the overall design goals.

Technically, we compose this reduced set of power quantities to be minimized, by combining together constraints on adjacent meta-atoms; for $K$ meta-atoms, we need to reduce the $2K-1$ constraints of \eqref{eq:passivity_condition} and \eqref{eq:lossless_condition} to fit the $K-1$ degrees of freedom corresponding to the horizontal offsets of the meta-atoms $d_2$, $d_3$, ..., $d_K$ with respect to the origin. Specifically, for the case of the $K=4$ prototype implementing perfect anomalous refraction (Section \ref{subsec:refractor}), the quantities to be minimized are chosen as $(a_1+a_2)/2$ , $(a_3+a_4)/2$ and $(b1+b_2+b_3+b_4)/4$. In other words, we require, respectively, that the load specifications of meta-atoms 1 and 2 combined to be as reactive as possible; the load specifications of meta-atoms 3 and 4 combined to be as reactive as possible; and the overall dissipated power in all of the 4 meta-atoms together to be minimal. As each one of the power quantities $a_k$, $b_k$ is non-negative, their average is non-negative, and the averaged quantities can be safely used as equivalent constraints for minimization, without damaging, formally, the design goal. This procedure yields 3 constraints for the anomalous refraction MG, matching the number of available degree of freedom \emph{in practice}, $d_2$ ,$d_3$, and $d_4$, enabling execution of \texttt{lsqnonlin}.

To use \texttt{lsqnonlin} for searching the optimal meta-atom distribution $d_2$ ,$d_3$, and $d_4$ in view of these nonlinear constraints, one has to provide initial values for the optimization process. 
As we have no preferred choice for these parameters, we generate the initial values randomly. We repeat the optimization process $50$ times (overall runtime $<5\mathrm{min}$) with different random initial values, and choose out of the resulting $50$ sets of MG designs the one best matching the passivity and loss constraints (the lowest residual values of $a_k$ and $b_k$); for the anomslous refraction prototype, these are listed in Table \ref{tab:Refractor}. 

Similarly, for the non-local lens array MG with $K=12$ meta-atoms (Section \ref{subsec:lens}), the fixed laminate thicknesses implied that the $2K-1=23$ nonlinear constraints formulated in \eqref{eq:passivity_condition} and \eqref{eq:lossless_condition} should be compacted to fit the $K-1=11$ available degrees of freedom $d_2,...,d_{12}$. As an extension of the compacting methodology we followed for the anomalous refraction MG prototype, the chosen power quantities to minimize were formed by averaging every two consecutive parameters as much as possible, namely, $(a_k+a_{k+1})/2$ for $k=1,3,5,7,9,11$, $(b_k+b_{k+1})/2$ for $k=1,3,5,7$, and $(b_9+b_{10}+b_{11}+b_{12})/4$, yielding $11$ constraints, as required. Again, the design specifications listed in Table \ref{tab:Lens} were obtained by running \texttt{lsqnonlin} for 50 times with random initial values for $d_2,...,d_{12}$ and choosing the meta-atom distribution with the least deviation from the prescribed constraints.

It should be noted that more sophisticated optimization methods can be harnessed. In particular, one can significantly increase the effective number of available degrees of freedom by considering a discrete set of commercially available laminate thicknesses and permittivities [note that our analytical model (Section \ref{sec:Theory}) can accommodate any stack of dielectrics], and allow the optimization algorithm to choose the best stratified media composition and even the distribution of meta-atoms between layers. In this paper, our aim was solely to demonstrate the effectiveness of the proposed synthesis method in obtaining detailed multilayer PCB layouts for MGs implementing versatile diffraction engineering without full-wave optimization; hence, we chose simple algorithms, which, as shown, were sufficient to this end.


\begin{thebibliography}{10}
\providecommand{\url}[1]{#1}
\csname url@samestyle\endcsname
\providecommand{\newblock}{\relax}
\providecommand{\bibinfo}[2]{#2}
\providecommand{\BIBentrySTDinterwordspacing}{\spaceskip=0pt\relax}
\providecommand{\BIBentryALTinterwordstretchfactor}{4}
\providecommand{\BIBentryALTinterwordspacing}{\spaceskip=\fontdimen2\font plus
\BIBentryALTinterwordstretchfactor\fontdimen3\font minus
  \fontdimen4\font\relax}
\providecommand{\BIBforeignlanguage}[2]{{%
\expandafter\ifx\csname l@#1\endcsname\relax
\typeout{** WARNING: IEEEtran.bst: No hyphenation pattern has been}%
\typeout{** loaded for the language `#1'. Using the pattern for}%
\typeout{** the default language instead.}%
\else
\language=\csname l@#1\endcsname
\fi
#2}}
\providecommand{\BIBdecl}{\relax}
\BIBdecl

\bibitem{born1999principles}
M.~Born and E.~Wolf, \emph{Principles of Optics}.\hskip 1em plus 0.5em minus
  0.4em\relax New York: Cambridge University Press, 1999.

\bibitem{goodman2005introduction}
J.~W. Goodman, \emph{{Introduction to Fourier Optics}}.\hskip 1em plus 0.5em
  minus 0.4em\relax Roberts and Company Publishers, 2005.

\bibitem{yariv2006photonics}
A.~Yariv and P.~Yeh, \emph{{Photonics: Optical Electronics in Modern
  Communications}}.\hskip 1em plus 0.5em minus 0.4em\relax Oxford University
  Press, 2006.

\bibitem{balanis2005antenna}
C.~A. Balanis, \emph{Antenna Theory: Analysis and Design}.\hskip 1em plus 0.5em
  minus 0.4em\relax Hoboken, NJ: {John wiley \& sons}, 2005.

\bibitem{balanis2012advanced}
------, \emph{Advanced Engineering Electromagnetics}.\hskip 1em plus 0.5em
  minus 0.4em\relax Hoboken, NJ: {John Wiley \& Sons}, 2012.

\bibitem{knott2004radar}
E.~F. Knott, J.~F. Schaeffer, and M.~T. Tulley, \emph{Radar Cross
  Section}.\hskip 1em plus 0.5em minus 0.4em\relax Raleigh, NC: SciTech
  Publishing, 2004.

\bibitem{fano1941theory}
U.~Fano, ``The theory of anomalous diffraction gratings and of quasi-stationary
  waves on metallic surfaces ({S}ommerfeld’s waves),'' \emph{J. Opt. Soc.
  Am.}, vol.~31, no.~3, pp. 213--222, 1941.

\bibitem{loewen1997diffraction}
E.~G. Loewen and E.~Popov, \emph{Diffraction Gratings and Applications}.\hskip
  1em plus 0.5em minus 0.4em\relax CRC Press, 1997.

\bibitem{perry1995high}
M.~Perry, C.~Shannon, E.~Shults, R.~Boyd, J.~Britten, D.~Decker, and B.~Shore,
  ``High-efficiency multilayer dielectric diffraction gratings,'' \emph{Opt.
  Lett.}, vol.~20, no.~8, pp. 940--942, 1995.

\bibitem{hehl1999high}
K.~Hehl, J.~Bischoff, U.~Mohaupt, M.~Palme, B.~Schnabel, L.~Wenke,
  R.~B\"{o}defeld, W.~Theobald, E.~Welsch, R.~Sauerbrey, and H.~Heyer,
  ``High-efficiency dielectric reflection gratings: design, fabrication, and
  analysis,'' \emph{Appl. Opt.}, vol.~38, no.~30, pp. 6257--6271, Oct 1999.

\bibitem{felsen1994radiation}
L.~B. Felsen and N.~Marcuvitz, \emph{Radiation and Scattering of Waves}.\hskip
  1em plus 0.5em minus 0.4em\relax Hoboken, NJ: John Wiley \& Sons, 1994.

\bibitem{bomzon2001pancharatnam}
Z.~Bomzon, V.~Kleiner, and E.~Hasman, ``Pancharatnam--{B}erry phase in
  space-variant polarization-state manipulations with subwavelength gratings,''
  \emph{Opt. Lett.}, vol.~26, no.~18, pp. 1424--1426, 2001.

\bibitem{yu2011light}
N.~Yu, P.~Genevet, M.~A. Kats, F.~Aieta, J.-P. Tetienne, F.~Capasso, and
  Z.~Gaburro, ``Light propagation with phase discontinuities: generalized laws
  of reflection and refraction,'' \emph{Science}, vol. 334, no. 6054, pp.
  333--337, 2011.

\bibitem{pfeiffer2013metamaterial}
C.~Pfeiffer and A.~Grbic, ``Metamaterial {Huygens'} surfaces: tailoring wave
  fronts with reflectionless sheets,'' \emph{Phys. Rev. Lett.}, vol. 110,
  no.~19, p. 197401, 2013.

\bibitem{monticone2013full}
F.~Monticone, N.~M. Estakhri, and A.~Al{\`u}, ``Full control of nanoscale
  optical transmission with a composite metascreen,'' \emph{Phys. Rev. Lett.},
  vol. 110, no.~20, p. 203903, 2013.

\bibitem{selvanayagam2013discontinuous}
M.~Selvanayagam and G.~V. Eleftheriades, ``Discontinuous electromagnetic fields
  using orthogonal electric and magnetic currents for wavefront manipulation,''
  \emph{Opt. Express}, vol.~21, no.~12, pp. 14\,409--14\,429, June 2013.

\bibitem{pfeiffer2013millimeter}
C.~Pfeiffer and A.~Grbic, ``Millimeter-wave transmitarrays for wavefront and
  polarization control,'' \emph{IEEE Trans. Microwave Theory Techn.}, vol.~61,
  no.~12, pp. 4407--4417, dec 2013.

\bibitem{pfeiffer2014efficient}
C.~Pfeiffer, N.~K. Emani, A.~M. Shaltout, A.~Boltasseva, V.~M. Shalaev, and
  A.~Grbic, ``Efficient light bending with isotropic metamaterial {H}uygens'
  surfaces,'' \emph{Nano Lett.}, vol.~14, pp. 2491--2497, apr 2014.

\bibitem{kim2014optical}
M.~Kim, A.~M.~H. Wong, and G.~V. Eleftheriades, ``Optical {H}uygens'
  metasurfaces with arbitrarily tailored local reflection coefficients,''
  \emph{Phys. Rev. X}, vol.~4, p. 041042, 2014.

\bibitem{asadchy2015functional}
V.~S. Asadchy, Y.~Ra’di, J.~Vehmas, and S.~A. Tretyakov, ``Functional
  metamirrors using bianisotropic elements,'' \emph{Phys. Rev. Lett.}, vol.
  114, no.~9, p. 095503, 2015.

\bibitem{epstein2016huygens}
A.~Epstein and G.~V. Eleftheriades, ``Huygens’ metasurfaces via the
  equivalence principle: design and applications,'' \emph{J. Opt. Soc. Am. B},
  vol.~33, no.~2, pp. A31--A50, 2016.

\bibitem{estakhri2016recent}
N.~M. Estakhri and A.~Al{\`u}, ``Recent progress in gradient metasurfaces,''
  \emph{J. Opt. Soc. Am. B}, vol.~33, no.~2, pp. A21--A30, 2016.

\bibitem{tretyakov2003analytical}
S.~Tretyakov, \emph{Analytical Modeling in Applied Electromagnetics}.\hskip 1em
  plus 0.5em minus 0.4em\relax Artech House, 2003.

\bibitem{kuester2003averaged}
E.~F. Kuester, M.~A. Mohamed, M.~Piket-May, and C.~L. Holloway, ``Averaged
  transition conditions for electromagnetic fields at a metafilm,'' \emph{IEEE
  Trans. Antennas Propag.}, vol.~51, no.~10, pp. 2641--2651, 2003.

\bibitem{glybovski2016metasurfaces}
S.~B. Glybovski, S.~A. Tretyakov, P.~A. Belov, Y.~S. Kivshar, and C.~R.
  Simovski, ``Metasurfaces: From microwaves to visible,'' \emph{Phys. Rep.},
  vol. 634, pp. 1--72, 2016.

\bibitem{wong2016reflectionless}
J.~P.~S. Wong, A.~Epstein, and G.~V. Eleftheriades, ``Reflectionless wide-angle
  refracting metasurfaces,'' \emph{IEEE Antennas Wireless Propag. Lett.},
  vol.~15, pp. 1293--1296, 2016.

\bibitem{epstein2016arbitrary}
A.~Epstein and G.~V. Eleftheriades, ``Arbitrary power-conserving field
  transformations with passive lossless omega-type bianisotropic
  metasurfaces,'' \emph{IEEE Trans. Antennas Propag.}, vol.~64, no.~9, pp.
  3880--3895, 2016.

\bibitem{asadchy2016perfect}
V.~S. Asadchy, M.~Albooyeh, S.~N. Tcvetkova, A.~D{\'\i}az-Rubio, Y.~Ra'di, and
  S.~A. Tretyakov, ``Perfect control of reflection and refraction using
  spatially dispersive metasurfaces,'' \emph{Phys. Rev. B}, vol.~94, no.~7, p.
  075142, 2016.

\bibitem{epstein2016synthesis}
A.~Epstein and G.~V. Eleftheriades, ``Synthesis of passive lossless
  metasurfaces using auxiliary fields for reflectionless beam splitting and
  perfect reflection,'' \emph{Phys. Rev. Lett.}, vol. 117, no.~25, p. 256103,
  2016.

\bibitem{kwon2017perfect}
D.~H. Kwon and S.~A. Tretyakov, ``{Perfect reflection control for impenetrable
  surfaces using surface waves of orthogonal polarization},'' \emph{Phys. Rev.
  B}, vol.~96, no.~8, p. 085438, aug 2017.

\bibitem{pfeiffer2014bianisotropic}
C.~Pfeiffer and A.~Grbic, ``Bianisotropic metasurfaces for optimal polarization
  control: Analysis and synthesis,'' \emph{Phys. Rev. Appl.}, vol.~2, no.~4, p.
  044011, 2014.

\bibitem{epstein2016cavity}
A.~Epstein, J.~P.~S. Wong, and G.~V. Eleftheriades, ``Cavity-excited
  {Huygens’} metasurface antennas for near-unity aperture illumination
  efficiency from arbitrarily large apertures,'' \emph{Nat. Commun.}, vol.~7,
  p. 10360, 2016.

\bibitem{lavigne2018susceptibility}
G.~Lavigne, K.~Achouri, V.~S. Asadchy, S.~A. Tretyakov, and C.~Caloz,
  ``Susceptibility derivation and experimental demonstration of refracting
  metasurfaces without spurious diffraction,'' \emph{IEEE Trans. Antennas
  Propag.}, vol.~66, no.~3, pp. 1321--1330, 2018.

\bibitem{chen2018theory}
M.~Chen, E.~Abdo-S{\'a}nchez, A.~Epstein, and G.~V. Eleftheriades, ``Theory,
  design, and experimental verification of a reflectionless bianisotropic
  {Huygens'} metasurface for wide-angle refraction,'' \emph{Phys. Rev. B},
  vol.~97, no.~12, p. 125433, 2018.

\bibitem{estakhri2016wave}
N.~M. Estakhri and A.~Al{\`u}, ``Wave-front transformation with gradient
  metasurfaces,'' \emph{Phys. Rev. X}, vol.~6, no.~4, p. 041008, 2016.

\bibitem{tretyakov2015metasurfaces}
S.~A. Tretyakov, ``Metasurfaces for general transformations of electromagnetic
  fields,'' \emph{Phil. Trans. R. Soc. A}, vol. 373, no. 2049, 2015.

\bibitem{epstein2017arbitrary}
A.~Epstein and G.~V. Eleftheriades, ``Arbitrary antenna arrays without feed
  networks based on cavity-excited omega-bianisotropic metasurfaces,''
  \emph{IEEE Trans. Antennas Propag.}, vol.~65, no.~4, pp. 1749--1756, 2017.

\bibitem{kwon2018lossless}
D.-H. Kwon, ``Lossless scalar metasurfaces for anomalous reflection based on
  efficient surface field optimization,'' \emph{IEEE Antennas Wireless Propag.
  Lett.}, vol.~17, no.~7, pp. 1149--1152, 2018.

\bibitem{kwon2018lossless1}
------, ``Lossless tensor surface electromagnetic cloaking for large objects in
  free space,'' \emph{Phys. Rev. B}, vol.~98, p. 125137, Sep 2018.

\bibitem{dorrah2018bianisotropic}
A.~H. {Dorrah} and G.~V. {Eleftheriades}, ``Bianisotropic huygens’
  metasurface pairs for nonlocal power-conserving wave transformations,''
  \emph{IEEE Antenna Wireless Propag. Lett.}, vol.~17, no.~10, pp. 1788--1792,
  Oct 2018.

\bibitem{raeker2019compound}
B.~O. Raeker and A.~Grbic, ``Compound metaoptics for amplitude and phase
  control of wave fronts,'' \emph{Phys. Rev. Lett.}, vol. 122, p. 113901, Mar
  2019.

\bibitem{ra2017meta}
Y.~Ra'di, D.~L. Sounas, and A.~Al{\`{u}}, ``Metagratings: {Beyond} the limits
  of graded metasurfaces for wave front control,'' \emph{Phys. Rev. Lett.},
  vol. 119, no.~6, p. 067404, aug 2017.

\bibitem{chalabi2017efficient}
H.~Chalabi, Y.~Ra'di, D.~L. Sounas, and A.~Al{\`u}, ``Efficient anomalous
  reflection through near-field interactions in metasurfaces,'' \emph{Phys.
  Rev. B}, vol.~96, no.~7, p. 075432, 2017.

\bibitem{epstein2017unveiling}
A.~Epstein and O.~Rabinovich, ``Unveiling the properties of metagratings via a
  detailed analytical model for synthesis and analysis,'' \emph{Phys. Rev.
  Appl.}, vol.~8, p. 054037, Nov 2017.

\bibitem{rabinovich2018analytical}
O.~Rabinovich and A.~Epstein, ``Analytical design of printed-circuit-board
  {(PCB)} metagratings for perfect anomalous reflection,'' \emph{IEEE Trans.
  Antennas Propag.}, vol.~66, no.~8, pp. 4086--4095, 2018.

\bibitem{rabinovich2019experimental}
O.~Rabinovich, I.~Kaplon, J.~Reis, and A.~Epstein, ``Experimental demonstration
  and in-depth investigation of analytically designed anomalous reflection
  metagratings,'' \emph{Physical Review B}, vol.~99, no.~12, p. 125101, 2019.

\bibitem{wong2018perfect}
A.~M.~H. Wong and G.~V. Eleftheriades, ``Perfect anomalous reflection with a
  bipartite {Huygens'} metasurface,'' \emph{Phys. Rev. X}, vol.~8, p. 011036,
  Feb 2018.

\bibitem{khaidarov2017asymmetric}
E.~Khaidarov, H.~Hao, R.~Paniagua-Dom{\'\i}nguez, Y.~F. Yu, Y.~H. Fu,
  V.~Valuckas, S.~L.~K. Yap, Y.~T. Toh, J.~S.~K. Ng, and A.~I. Kuznetsov,
  ``Asymmetric nanoantennas for ultrahigh angle broadband visible light
  bending,'' \emph{Nano Lett.}, vol.~17, no.~10, pp. 6267--6272, 2017.

\bibitem{yang2017freeform}
J.~Yang, D.~Sell, and J.~A. Fan, ``Freeform metagratings based on complex light
  scattering dynamics for extreme, high efficiency beam steering,''
  \emph{Annal. Phys.}, p. 1700302, oct 2017.

\bibitem{paniagua2018metalens}
R.~Paniagua-Domínguez, Y.~F. Yu, E.~Khaidarov, S.~Choi, V.~Leong, R.~M.
  Bakker, X.~Liang, Y.~H. Fu, V.~Valuckas, L.~A. Krivitsky, and A.~I.
  Kuznetsov, ``A metalens with a near-unity numerical aperture,'' \emph{Nano
  Lett.}, vol.~18, no.~3, pp. 2124--2132, 2018.

\bibitem{sell2018ultra}
D.~Sell, J.~Yang, E.~W. Wang, T.~Phan, S.~Doshay, and J.~A. Fan,
  ``Ultra-high-efficiency anomalous refraction with dielectric metasurfaces,''
  \emph{ACS Photonics}, vol.~5, no.~6, pp. 2402--2407, jun 2018.

\bibitem{fan2018perfect}
Z.~Fan, M.~R. Shcherbakov, M.~Allen, J.~Allen, B.~Wenner, and G.~Shvets,
  ``Perfect diffraction with multiresonant bianisotropic metagratings,''
  \emph{ACS Photonics}, vol.~5, no.~11, pp. 4303--4311, 2018.

\bibitem{wong2018binary}
A.~M.~H. Wong, P.~Christian, and G.~V. Eleftheriades, ``Binary {Huygens'}
  metasurfaces: {Experimental} demonstration of simple and efficient
  near-grazing retroreflectors for {TE and TM} polarizations,'' \emph{IEEE
  Trans. Antennas Propag.}, vol.~66, no.~6, pp. 2892--2903, jun 2018.

\bibitem{neder2019combined}
V.~Neder, Y.~Ra'di, A.~Al{\`u}, and A.~Polman, ``Combined metagratings for
  efficient broad-angle scattering metasurface,'' \emph{ACS Photonics}, 2019.

\bibitem{popov2018controlling}
V.~Popov, F.~Boust, and S.~N. Burokur, ``Controlling diffraction patterns with
  metagratings,'' \emph{Phys. Rev. Appl.}, vol.~10, no.~1, p. 011002, 2018.

\bibitem{popov2019constructing}
------, ``Constructing the near field and far field with reactive metagratings:
  Study on the degrees of freedom,'' \emph{Phys. Rev. Appl.}, vol.~11, no.~2,
  p. 024074, 2019.

\bibitem{popov2019designing}
V.~Popov, M.~Yakovleva, F.~Boust, J.-L. Pelouard, F.~Pardo, and S.~N. Burokur,
  ``Designing metagratings via local periodic approximation: From microwaves to
  infrared,'' \emph{Phys. Rev. Appl.}, vol.~11, p. 044054, Apr 2019.

\bibitem{epstein2018eucap}
A.~Epstein and O.~Rabinovich, ``Perfect anomalous refraction with
  metagratings,'' in \emph{Proc. 12th European Conf. Antennas and Propagation
  (EUCAP)}, London, UK, 2018.

\bibitem{packo2019inverse}
P.~Packo, A.~N. Norris, and D.~Torrent, ``Inverse grating problem: Efficient
  design of anomalous flexural wave reflectors and refractors,'' \emph{Phys.
  Rev. Appl.}, vol.~11, no.~1, p. 014023, 2019.

\bibitem{ikonen2007modeling}
P.~M.~T. Ikonen, E.~Saenz, R.~Gonzalo, and S.~A. Tretyakov, ``{Modeling and
  analysis of composite antenna superstrates consisting on grids of loaded
  wires},'' \emph{IEEE Trans. Antennas Propag.}, vol.~55, no.~10, pp.
  2692--2700, 2007.

\bibitem{Chew1990}
W.~C. Chew, \emph{{Waves and Fields in Inhomogeneous Media}}.\hskip 1em plus
  0.5em minus 0.4em\relax New York: Van Nostrand Reinhold, 1990.

\bibitem{osipov2017modern}
A.~V. Osipov and S.~A. Tretyakov, \emph{Modern Electromagnetic Scattering
  Theory with Applications}.\hskip 1em plus 0.5em minus 0.4em\relax John Wiley
  \& Sons, 2017.

\bibitem{epstein2010impact}
A.~Epstein, N.~Tessler, and P.~D. Einziger, ``The impact of spectral and
  spatial exciton distributions on optical emission from thin-film
  weak-microcavity organic light-emitting diodes,'' \emph{IEEE J. Quantum
  Electron.}, vol.~46, no.~9, pp. 1388--1395, 2010.

\bibitem{epstein2013thesis}
\BIBentryALTinterwordspacing
A.~Epstein, ``Rigorous electromagnetic analysis of optical emission of organic
  light-emitting diodes,'' Ph.D. dissertation, Dept. Elect. Eng., Technion -
  Israel Insititute of Technology, Haifa, Israel, 2013. [Online]. Available:
  \url{http://arielepstein.webs.com/Research/thesis.pdf}
\BIBentrySTDinterwordspacing

\bibitem{guptamicrostrip}
K.~C. Gupta, R.~Garg, I.~Bahl, and P.~Bhartia, \emph{Microstrip Lines and
  Slotlines}.\hskip 1em plus 0.5em minus 0.4em\relax Artech House, 1996.

\bibitem{diaz2017generalized}
A.~D{\'\i}az-Rubio, V.~S. Asadchy, A.~Elsakka, and S.~A. Tretyakov, ``From the
  generalized reflection law to the realization of perfect anomalous
  reflectors,'' \emph{Science Adv.}, vol.~3, no.~8, p. e1602714, 2017.

\bibitem{asadchy2017flat}
V.~S. Asadchy, A.~D{\'\i}az-Rubio, S.~N. Tcvetkova, D.-H. Kwon, A.~Elsakka,
  M.~Albooyeh, and S.~A. Tretyakov, ``Flat engineered multichannel
  reflectors,'' \emph{Phys. Rev. X}, vol.~7, no.~3, p. 031046, 2017.

\bibitem{hasman2003polarization}
E.~Hasman, V.~Kleiner, G.~Biener, and A.~Niv, ``{Polarization dependent
  focusing lens by use of quantized Pancharatnam-Berry phase diffractive
  optics},'' \emph{Appl. Phys. Lett.}, vol.~82, no.~3, pp. 328--330, 2003.

\bibitem{lin2014dielectric}
D.~Lin, P.~Fan, E.~Hasman, and M.~L. Brongersma, ``Dielectric gradient
  metasurface optical elements,'' \emph{Science}, vol. 345, no. 6194, pp.
  298--302, 2014.

\bibitem{aieta2015multiwavelength}
F.~Aieta, M.~A. Kats, P.~Genevet, and F.~Capasso, ``Multiwavelength achromatic
  metasurfaces by dispersive phase compensation,'' \emph{Science}, vol. 347,
  no. 6228, pp. 1342--1345, 2015.

\bibitem{chen2018broadband}
W.~T. Chen, A.~Y. Zhu, V.~Sanjeev, M.~Khorasaninejad, Z.~Shi, E.~Lee, and
  F.~Capasso, ``A broadband achromatic metalens for focusing and imaging in the
  visible,'' \emph{Nature Nanotechnol.}, vol.~13, no.~3, p. 220, 2018.

\end{thebibliography}
\end{document}